\documentclass[sigplan,screen]{acmart}

\usepackage[]{hyperref}

\newcommand\ket[1]{\left|#1\right\rangle}

\newcommand{\highwaycolor}{dark blue}

\usepackage{amsmath,mathtools}

\usepackage{listings}
\usepackage{multirow}
\usepackage{multicol}
\usepackage{url}
\usepackage[ruled,vlined,linesnumbered]{algorithm2e}
\usepackage{tikz}
\usepackage{xcolor}
\usepackage{ctable} 
\usepackage{stfloats}

\newcommand{\frameworkname}{our framework}

\copyrightyear{2024} 
\acmYear{2024} 
\setcopyright{rightsretained} 
\acmConference[ASPLOS '24]{29th ACM International Conference on Architectural Support for Programming Languages and Operating Systems, Volume 2}{April 27-May 1, 2024}{La Jolla, CA, USA}
\acmBooktitle{29th ACM International Conference on Architectural Support for Programming Languages and Operating Systems, Volume 2 (ASPLOS '24), April 27-May 1, 2024, La Jolla, CA, USA}
\acmDOI{10.1145/3620665.3640377}
\acmISBN{979-8-4007-0385-0/24/04}
\begin{document}

\title{MECH: Multi-Entry Communication Highway for Superconducting Quantum Chiplets}

\author{Hezi Zhang}
\email{hezi@ucsd.edu}
\affiliation{%
  \institution{University of California,}
  \city{San Diego}
  \country{USA}
}
\author{Keyi Yin}
\email{keyin@ucsd.edu}
\affiliation{%
  \institution{University of California,}
  \city{San Diego}
  \country{USA}
}
\author{Anbang Wu}
\email{anbang@cs.ucsb.edu}
\affiliation{%
  \institution{University of California,}
  \city{Santa Barbara}
  \country{USA}
}
\author{Hassan Shapourian}
\email{hshapour@cisco.com}
\affiliation{%
  \institution{Cisco Quantum Lab}
  \city{San Jose}
  \country{USA}
}
\author{Alireza Shabani}
\email{ashabani@cisco.com}
\affiliation{%
  \institution{Cisco Quantum Lab}
  \city{Santa Monica}
  \country{USA}
}
\author{Yufei Ding}
\email{yufeiding@ucsd.edu}
\affiliation{%
  \institution{University of California,}
  \city{San Diego}
  \country{USA}
}

\renewcommand{\shortauthors}{H. Zhang, K. Yin, A. Wu, H. Shapourian, A. Shabani, Y. Ding}

\begin{abstract}

Chiplet architecture is an emerging architecture for quantum computing that could significantly increase qubit resources with its great scalability and modularity. However, as the computing scale increases, communication between qubits would become a more severe bottleneck due to the long routing distances. In this paper, we propose a multi-entry communication highway (MECH) mechanism to trade ancillary qubits for program concurrency, and build a compilation framework to efficiently manage and utilize the highway resources.
Our evaluation shows that this framework significantly outperforms the baseline approach in both the circuit depth and the number of operations on typical quantum benchmarks. This implies a more efficient and 
less error-prone compilation of quantum programs.

\end{abstract}

\maketitle 

\section{Introduction}
The past decade has witnessed exciting breakthroughs of quantum computing technologies on various hardware platforms~\cite{hardware_SC_1_supremacy,hardware_SC_2,hardware_SC_3_supremacy,hardware_TI,hardware_NI,hardware_photon_1_supremacy,hardware_photon_2}, with the superconducting platform being one of the leading candidates.
However, the complexity of building increasingly larger devices \cite{larger_devices_1, larger_devices_2} and the capacity limitation of individual cryogenic dilution refrigerators~\cite{refrigerator_capacity, multi-node} bring extreme challenges for eventually realizing a giant superconducting quantum chip.
Moreover, the fabrication yields will also decline as the number of qubits on the chip increases, leading to a significant increase in manufacturing costs.
To realize large-scale quantum computers, a modular system linking processors together, either with long-range or short-range links~\cite{DQN_hardware_1,DQC_hardware_2,DQC_hardware_3,DQC_hardware_4, DQC_hardware_5}, is required.

Recently, there has been significant research interest~\cite{chiplet_1,chiplet_2,chiplet_3, chiplet_4_latest_fred_chong} in the modular architecture that connects nearby small quantum processors, known as chiplets, due to advancements in short-range inter-chip connections on superconducting  platforms~\cite{short_range_hardware_1,short_range_hardware_2,short_range_hardware_3,short_range_hardware_4,short_range_hardware_5}.
These chiplets form a multi-chip processor that offers a middle ground between monolithic computing and long-range distributed quantum networks.
This approach is expected to enable 1-10 thousand qubits in the near term \cite{IBM2025roadmap}, with physical gate error rates ranging from 1e-4 (e.g. single-qubit gates \cite{99.77}) to 1e-2 (e.g. cross-chip links \cite{short_range_hardware_5}).
Indeed, it is essential to acknowledge that this computing scale remains a considerable distance from reaching the real-world fault-tolerant quantum computing, as the 1-10 thousand physical qubits could accommodate only several long-lived logical qubits \cite{google_blog} and the gate error rates are still above the fault-tolerant threshold \cite{surface_code}.
However, it already has the potential to significantly advance the capabilities of near-term quantum computing and
empower a range of applications in the NISQ era.

Despite the significant leap in scale, this emerging chiplet architecture presents a new set of challenges which requires novel techniques in compilation of quantum programs. 
First, with the increased computing scale comes longer execution time of quantum programs, which places a higher demand on the coherence time of the qubits. 
Second, the different technologies used for on-chip and cross-chip connections introduce heterogeneity in their characteristics, with the fidelity of cross-chip connections typically lower than that of on-chip connections. 
Third, in contrast to the possibility of all-to-all connectivity between processors in long-range distributed quantum computing, cross-chip connections in the chiplet architecture are subject to constraints that limit connectivity to neighboring chiplets only,
and these cross-chip connections may be sparser than on-chip ones.
These challenges cannot be addressed directly by existing compilers which are designed and tailored for either monolithic or (long-range) distributed quantum computing for the following reasons.

On one hand, compilation techniques for monolithic quantum computing is not efficient at the scale of computing on chiplets.
Monolithic compilers rely on insertion of SWAP gates to route qubits of each gate toward each other~\cite{Qiskit}, thus enabling the multi-qubit gate execution between distant qubits.
However, the routing paths of qubits increase with the scale of computing, and the qubits may route back and forth if they are involved in multiple gates.
Hence this approach would result in significant latency that soon becomes intolerable for the chiplet architecture.
Moreover, these compilers do not take into account the discrepancy between fidelities of on-chip and cross-chip connections.

On the other hand, compilation techniques for distributed quantum computing cannot be effectively applied to the chiplet architecture either.
Distributed compilers focus on minimizing the long-range communications, with the motivation that remote gates between different processors are much more expensive than the local ones within each processor.
However, this should not be the only objective for compilation on the chiplet architecture, as the disparity between on-chip connections and cross-chip connections is not as significant as that in the distributed quantum computing.
Furthermore, these compilers usually make some assumptions that are not applicable to the chiplet architecture, such as an all-to-all connectivity among different processors or easy access to the dedicated communication qubits from any data qubit.

In this paper, we investigate the optimization of quantum program compilation for the chiplet architecture to reveal the sweet spot between monolithic quantum computing and distributed quantum computing. 
%
%
The insight is that the aforementioned challenges can be overcome by innovative compilation techniques being novel at three levels.
%
First, the computing paradigm of the circuit model can be extended to trade additional  ancillary qubits for more concurrency of gate execution, thus reducing the execution time and mitigating the increasing demand for coherence time.
Second, the qubit communication can be boosted through an efficient communication mechanism over the ancillary qubits, which allocates ancillary qubits in a resource-efficient and architecture-aware manner to enable easy access to the ancillary resource while taking into account the discrepancy between on-chip and cross-chip links.
%
%
Third, gates in a program can be scheduled by a compiler to execute with or without the ancillary qubits based on their patterns in the program, with corresponding qubits routed accordingly by the compiler to respect the conncetivity constraints within or among the chiplets.
%

To this end, we propose a compilation framework that incorporates optimizations at the three levels above. 
\textbf{At the level of computing paradigm}, we incorporate ancillary qubits by adopting a hybrid of gate-based computing and measurement-based computing. In particular, the ancillary qubits operate in a measurement-based manner, taking advantage of the recent advancement of dynamic circuit~\cite{dynamic_circuit_1, dynamic_circuit_2}, while the regular data qubits remain in their purely gate-based manner.
This hybrid paradigm enables a tradeoff between additional qubit resources and program concurrency by leveraging the outstanding concurrency of the protocol in \cite{cat_entangle} (Fig.~\ref{fig:cat_TP}).
%
Besides the incorporation of measurements in the protocol itself, we further propose a measurement-based scheme for the entanglement preparation among the ancillary qubits, which facilitates the efficient realization of the protocol.

\textbf{At the level of communication mechanism}, we abstract the ancillary qubits into an adjustable \emph{multi-entry communication highway (MECH)}, providing a mechanism to efficiently utilize the ancillary resources with minimized qubit overhead. 
In particular, the highway is composed by consecutive paths of ancillary qubits, with the paths spanning across the chiplets to give easy access to all qubits.
The density of ancillary qubits along the paths are adaptive to the coupling structures of the chiplets, with sparse patterns allowed to reduce the qubit overhead and dense patterns enforced around critical positions to reduce the communication latency and address the architectural heterogeneity.
In this way, the highway forms a communication channel on the software level without the need of hardware modification, enabling concurrent gate execution regardless of the distances among the involved qubits.

\textbf{At the level of program compilation},
we fully exploit the highway channels with a dynamic scheduling of gates and an efficient routing of relevant qubits that maximizes the sharing of highway resources. 
In particular, the scheduling dynamically determines the periods of communication protocols based on the communication demand to accommodate as many gates as possible in each round, allowing a temporal sharing of highway resource among gates. The routing addresses the hardware connectivity constraints with optimizations from two aspects.
First, it routes necessary qubits towards nearby highway in a way that ensures the earliest execution of their gates. Second, it routes those gates through the highway in a way that minimizes routing path lengths by allowing spatial sharing of highway resources among qubits.
%
To summarize, our contributions in this paper are listed as follows:

\begin{itemize}

    \item We trade ancillary qubit resources for program concurrency by proposing a hybrid computing paradigm of gate-based and measurement-based computing, thus mitigating the enhanced demand for coherence time by larger scale of quantum computing.

    \item We abstract the ancillary qubits into an adjustable multi-entry communication highway at the software level, and propose a communication mechanism for allocation of highway with small overhead and efficient realization of the communication protocol.

    \item We perform compilation optimizations through a dynamic scheduling of gates onto the highway and an efficient routing of relevant qubits under the connectivity constraints, allowing for temporal and spatial sharing of highway to fully exploit the resources.

    
    \item We demonstrate a compilation framework that significantly outperforms the baseline in both the circuit depth and the number of operations. The evaluation shows a trend of reduced qubit overhead and increased outperformance as the computing scale increases, suggesting the scalibility of our framework.
\end{itemize}

\section{Background and Related Work}\label{sect:background}

\subsection{Quantum Computing Basics}

\begin{figure}[h]
    \centering
    \includegraphics[width=0.47\textwidth]{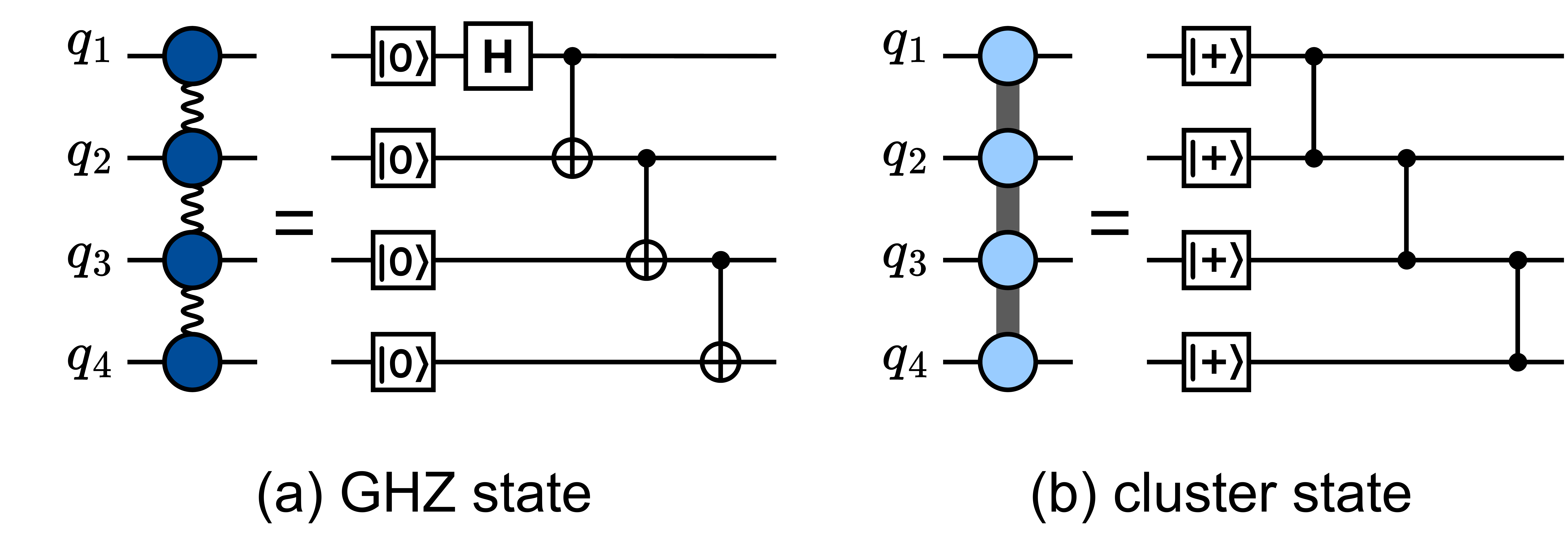}
    \caption{GHZ state and cluster state preparation.}
    \label{fig:Entangle}
\end{figure}

\paragraph{Entangled States}
Entangled states can serve as valuable resources for quantum computational tasks. For example, the \emph{GHZ states}
can be used to facilitate qubit communication over long distances. To prepare an N-qubit GHZ state, one can start with N qubits in the state $\left|0\right>$, and apply a chain of CNOT gates as illustrated in Fig.~\ref{fig:Entangle}(a), with the dark blue dots connected by wavy lines standing for the GHZ state.
Another example is the utilization of \emph{cluster states} \cite{mbqc2009}, a state of qubits arranged in a lattice $G=(V,E)$, enables universal computation in the measurement-based quantum computing (MBQC) model.
To prepare a cluster state, one can initialize all qubits in $G$ to the $\ket{+}$ state and then apply $CZ$ gates between each pair of neighboring qubits in graph $G$. A 1-dimensional example is illustrated in Fig.~\ref{fig:Entangle}(b), with the light blue dots connected by straight lines standing for the cluster state.

\begin{figure}[h]
    \centering
    \includegraphics[width=0.47\textwidth]{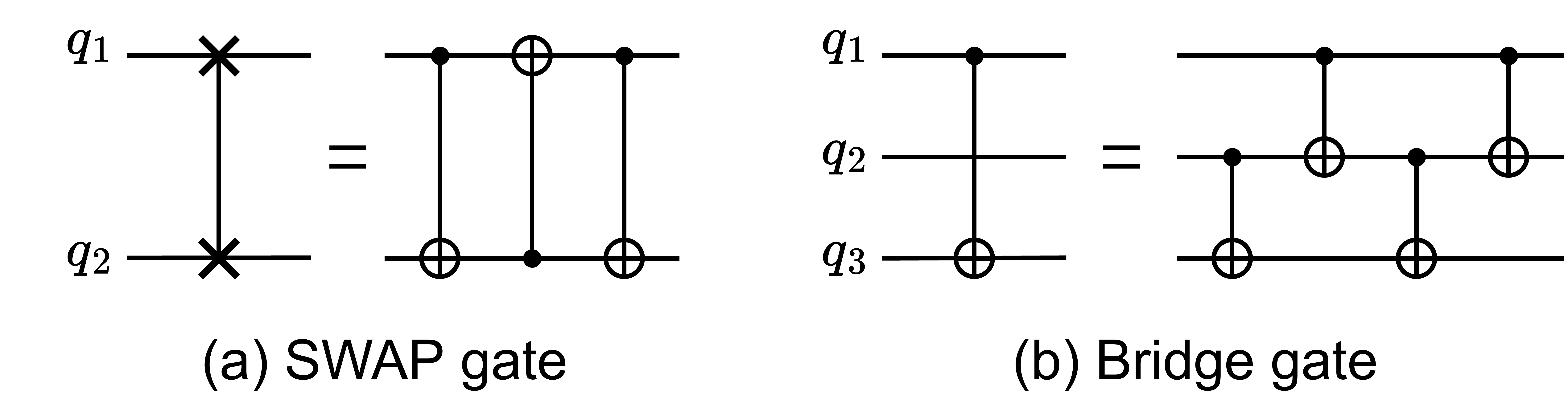}
    \caption{SWAP and bridge gates.}
    \label{fig:SWAP and Bridge}
\end{figure}

\paragraph{Quantum Gates}
In the circuit model of quantum computing, qubits are manipulated by quantum gates such as the 1-qubit Hadamard gate and 2-qubit CNOT gate. By arranging these basic gates appropriately, it is possible to abstract higher-level gates for communicating quantum data between qubits. For instance, a SWAP gate can exchange the states of two qubits, which can be achieved by 3 CNOT gates as illustrated in Fig.~\ref{fig:SWAP and Bridge}(a). A bridge gate can perform an effective CNOT  between two qubits that cannot interact with each other directly, through the use of a third qubit. This can be accomplished using 4 CNOT gates as depicted in Fig.~\ref{fig:SWAP and Bridge}(b). These gates play an important role in addressing the connectivity constraints between qubits on the hardware, and are widely used in various compilation frameworks.

\paragraph{Quantum Measurement}
Quantum measurement is a fundamental operation that allows us to extract information about the state of a quantum system. While measurements of quantum circuits are traditionally performed in a delay-to-the-end manner, recent advancement of hardware technologies~\cite{dynamic_circuit_1, dynamic_circuit_2} has enabled dynamic circuits with mid-circuit measurements.
This allows a seamless incorporation of real-time classical communication into quantum circuits, and will greatly increase the variety of circuits that can be executed on near-term quantum hardware, forming an important part of many quantum applications in the future.

\vspace{0.3em}
\subsection{Quantum Architecture and Compiler}

\paragraph{Monolithic Quantum Computing}
Monolithic architecture is the simplest architecture for quantum computing that presents a self-contained system on a single chip.
However, it faces extreme challenges in its scalability~\cite{chiplet_3}, as the increasing number of qubits would increase the noise, reduce the yield rate of manufacturing, and eventually push the capability of cooling and control technologies to their limit.

Compilation for monolithic quantum computing is responsible for transforming quantum circuits to a form that complies with the constraints of the underlying hardware. In particular, it addresses the connectivity constraints among physical qubits by routing logical qubits involved in the same gate to physical qubits that are directly connected on the hardware. On state-of-the-art compilers~\cite{Qiskit, staq, openql, tket}, this is achieved by insertion of SWAP or bridge gates into the circuit. However, as the scale of computing increases, this compilation approach will become inefficient, because the increasingly long routing paths will result in significant latency and harm the overall fidelity.

\paragraph{Distributed Quantum Networks}
A distributed architecture for quantum computing, also known as a distributed quantum network, comprises multiple processing nodes physically separated across a considerable distance. 
While quantum channels across the nodes can be realized by  microwave-to-optical (M2O) internode links~\cite{m2o1,m2o2,m2o3,m2o4,m2o5,m2o6}, these links are typically much noisier and slower than the connections within each node~\cite{DQN_hardware_1, chiplet_3}. 
Moreover, internode communications rely on complex protocols~\cite{2019NJPh, 2019arXiv190309778D}, 
usually including the preparation, distribution and measurement of EPR pairs. 
Consequently, remote operations between different nodes are considered much more expensive than the local ones.

\begin{figure}[h]
    \centering
    \includegraphics[width=0.31\textwidth]{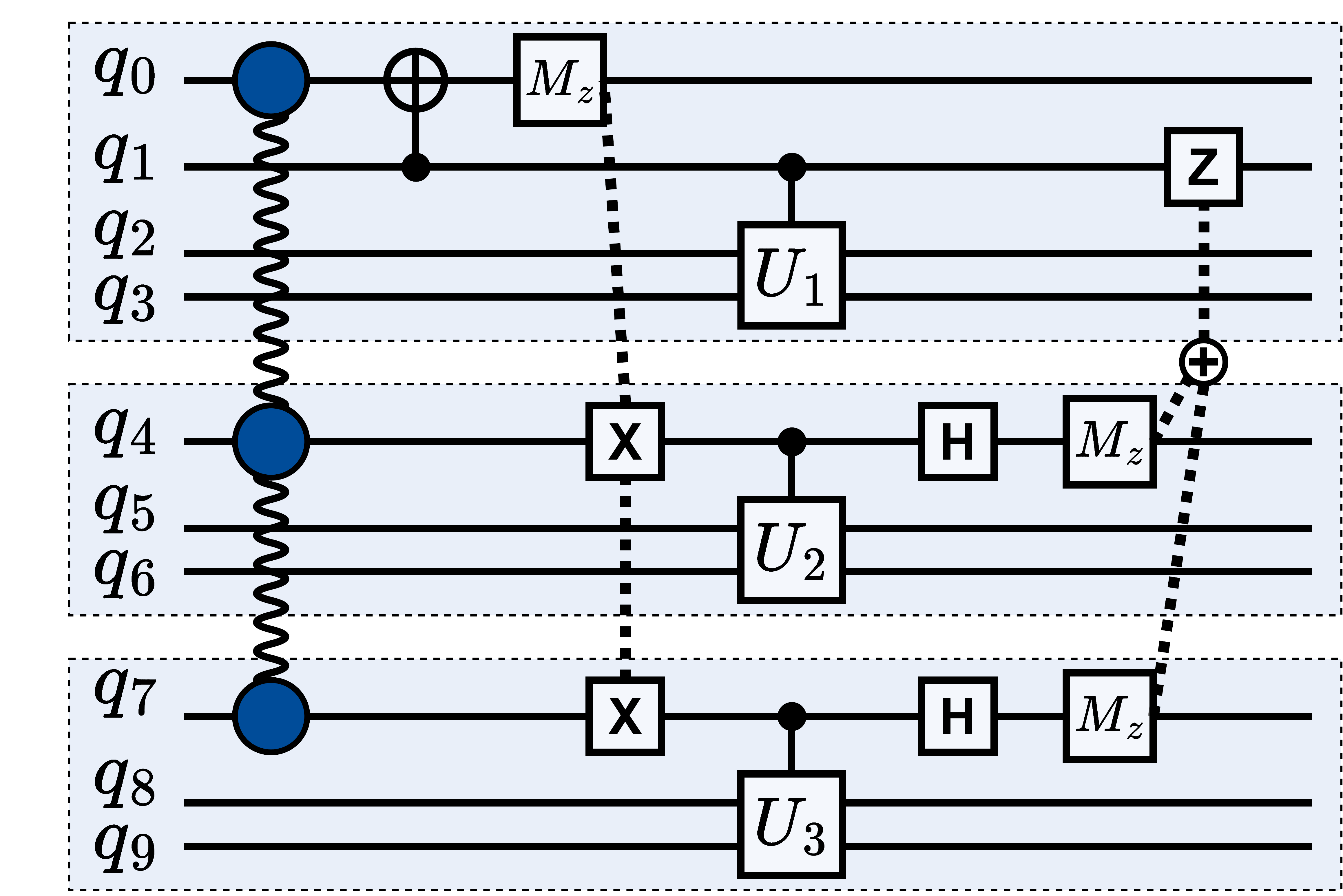}
    \caption{Communication protocol.}
    \label{fig:cat_TP}
\end{figure}

Compilers for the distributed architecture primarily focuses on minimizing the number of these expensive remote operations. 
This objective can be achieved by an efficient utilization of entangled states among different nodes. 
For example, \cite{cat_entangle} proposes a protocol that enables simultaneous data control on multiple nodes with the help of a pre-established GHZ state.
This protocol is depicted in Fig.~\ref{fig:cat_TP}, with the \highwaycolor\ nodes connected by wavy lines standing for the GHZ state. The three entangled qubits $q_0, q_4, q_7$ in the GHZ state are distributed to three different nodes, with the first node containing qubits $\{q_0, q_1, q_2, q_3\}$, the second node containing $\{q_4, q_5, q_6\}$ and the third one containing $\{q_7, q_8, q_9\}$. By entangling control qubit $q_1$ with a qubit $q_0$ in the GHZ state, conducting a measurement on $q_0$ with Pauli corrections on all the other qubits in the GHZ state, $q_1$ involves its data into the entangled state, and enables simultaneous control gates over qubits on all the three nodes.
However, in distributed compilers~\cite{dqn_compiler1,dqn_compiler2,dqn_compiler3,dqn_compiler4}, certain assumptions are usually made, such as that each node may communicate with any other node in the network, and the data qubits on each node can easily access the dedicated communication qubits on the same node.

\paragraph{Chiplet Architecture}
The chiplet architecture for quantum computing integrates multiple small, locally-connected chips, known as chiplets, to form a multi-chip module (MCM).
This can be achieved by linking separate physical modules to a larger carrier chip or interposer using flip-chip bonds
~\cite{short_range_hardware_5}, 
or connecting devices in different refrigerators using cryogenic links
~\cite{short_range_hardware_3}. The chiplet architecture offers greater scalability than the monolithic architecture, as the smaller size of the chiplets results in higher yield and lower fabrication costs, and reduces the complexity of design verification and post-fabrication testing. Furthermore, it has better performance than the distributed architecture, as the short-range cross-chip connections can be made with lower latency and higher fidelity compared to long-range internode links.

Recent research on the architecture side has shown the potential of chiplet architecture to exceed the performance of monolithic architecture on near-term hardware~\cite{chiplet_3} and demonstrated the feasibility of scaling quantum computing through modularity by analysis of yield and gate performance~\cite{chiplet_4_latest_fred_chong}.
However, 
there is still much work to be done on the compilation side.

\subsection{Communication between Qubits}
Quantum communication takes place via the movement of qubits or the transfer of their quantum states. 
\cite{ISCA03} introduces a quantum wire architecture based upon quantum teleportation to transport data over long distance, comparing it with the swapping channels by analyzing their latency, bandwidth, fabrication challenges with solid-state silicon technology and associated classical control logic.
\cite{ISCA06} investigates the construction of reliable long-distance quantum communication channels with ion trap technology by combining ballistic ion movement and quantum teleportation, addressing the routing problem for EPR pair distribution.
\cite{multipartite_photon} demonstrates on-chip genuine multipartite entanglement and quantum teleportation in silicon by coherently controlling an integrated network of microresonator nonlinear single-photon sources and linear optic multiqubit entangling circuits. \cite{multi_core_channel} proposes a multi-core architecture that utilizes dedicated communication qubits and specialized physical channels to perform inter-core quantum communication via quantum teleportation. \cite{teleport_mapping} incorporates the teleportation protocol in qubit mapping. But due to the lack of optimization, their compiler \cite{teleport_mapping_package} can only handle small-scale circuits, thus not applicable for the chiplet architecture.

Although quantum communication schemes over long distances have been described in the literature above, what has been missing is an efficient management of communication resources that allows different operations to share and utilize the available resources to their full potential. Our work fills this gap by conceptualizing the communication resources to an adjustable multi-entry communication highway (MECH). In this model, different qubits and gates can simultaneously access and share the resources, entering and exiting the highway at various points along it based on their needs. 
The efficient preparation and consumption of entanglement on the highway, and the precise management of the timing and trajectories of accessing the highway, all handled by our compiler, enhance the overall communication efficiency. Furthermore, as this highway channel is purely software-driven, it can be adjusted conveniently in response to the available qubits and to accommodate architectural heterogeneity.

\vspace{0.3em}
\section{Overview: Building Highways on Chiplets}

\begin{figure*}[t]
    \centering
    \includegraphics[width=0.97\textwidth]{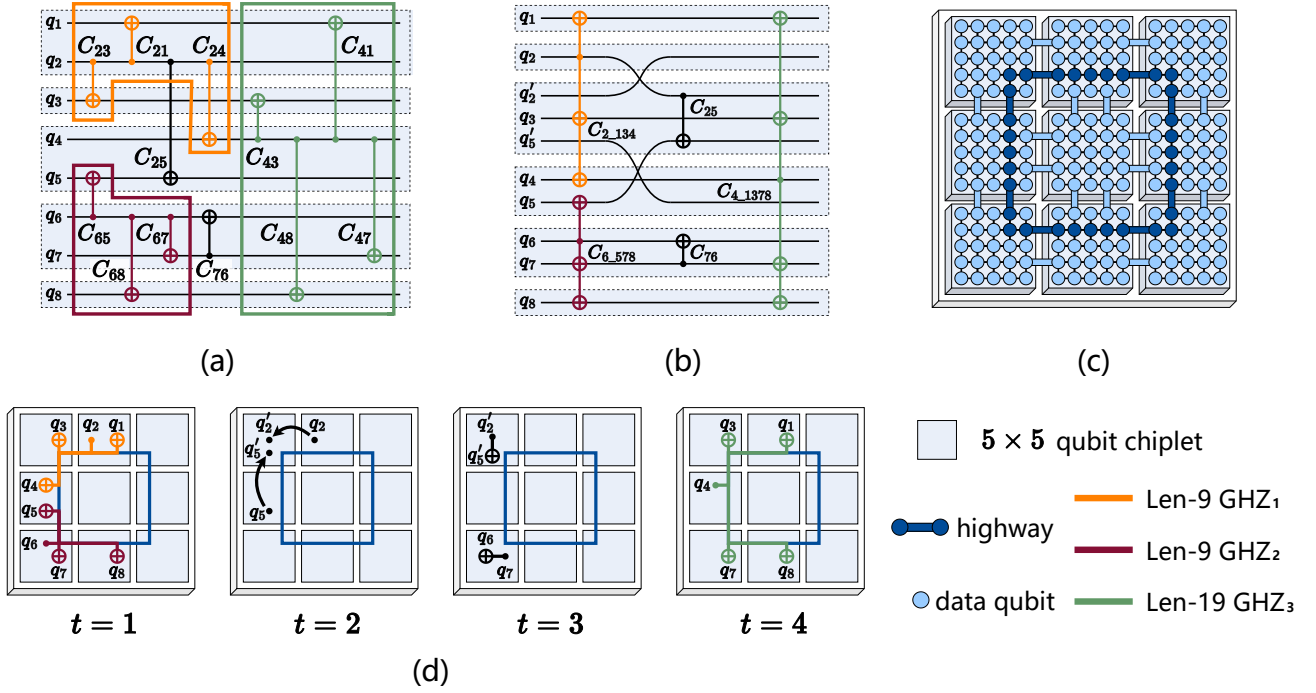}
    \caption{Computation process on chiplets in the presence of highway (\highwaycolor\ paths in (c)). Commutable gates in circuit (a) are aggregated to multi-target control gates (b) and executed simultaneously on the highway (d). }
    \label{fig:motivation_with_example}
\end{figure*}

As quantum computing scales up with the help of the emerging chiplet architecture, efficient communication between qubits would become a critical challenge. The increasing routing distances between qubits can lead to significant overhead in executing quantum programs, limiting the scale of programs that could be run within the coherence time. To overcome this challenge, a high-speed communication channel needs to be built on the chiplets to facilitate efficient communication between distant qubits, akin to building highways across cities to enable long-range transportation. 

Our insight is that this high-speed communication channel can be built on the software side, by enabling concurrency between gate execution regardless of the distances among the involved qubits. Specifically, the concurrency is achieved by resolving the following two problems. First, gate execution in the superconducting quantum computing system can not reach its maximum concurrency due to the hardware constraints. While in principle, execution of commutable gates in a circuit do not have dependency on each other, on the superconductive platform they have to be executed sequentially if they share common qubits. 
Second, communications between distant qubits are usually achieved through insertion of sequential SWAP gates between them. When qubits are involved in multiple gates, they often have to be routed back and forth, preventing the concurrent execution among those gates.

To build the communication channel, we provide an approach that allocates some of the qubits as ancillary qubits and establishes entanglements among them.
These entanglements can then be utilized as a computing resource to connect distant qubits and enhance the concurrency of their operations.
In particular, we utilize the communication protocol as shown in Fig.~\ref{fig:cat_TP} which allows simultaneous execution of controlled gates sharing the same control qubit.
%
Controlled gates sharing the same target qubit can also utilize this protocol by transforming themselves to controlled gates sharing the same control with Hadamard gates.
%
The entangled states required by the protocol are formed on the ancillary qubits, which are allocated close to each other on the hardware, forming a computational highway across chiplets.
By sacrificing a small portion of qubits, which are not as scarce as in monolithic computing with the highly scalable chiplet architecture, we can trade for more concurrency using this protocol originally proposed for distributed computing.

To illustrate the computation process in the presence of highway, Fig.~\ref{fig:motivation_with_example} shows an example of executing the circuit in Fig.~\ref{fig:motivation_with_example}(a) on a chiplet architecture in Fig.~\ref{fig:motivation_with_example}(c). 
This chiplet architecture consists of 3x3 chiplets each containing 5x5 qubits, with the highlighted qubits in \highwaycolor\ being the ancillary qubits that form a highway across the chiplets.
In the circuit (\ref{fig:motivation_with_example}(a)), gates $\left\{C_{21},C_{23},C_{24}\right\}$, gates $\left\{C_{65},C_{67},C_{68}\right\}$ and gates $\left\{C_{41},C_{43},C_{47},C_{48}\right\}$ 
are three sets of commutable gates that can be executed concurrently via the utilization of highway. 
Thus we aggregate them into multi-target controlled gates $C_{2\_134}, C_{6\_578}$ and $C_{4\_1378}$ (shown with corresponding colors in \ref{fig:motivation_with_example}(b)) while keeping the other gates (i.e., gates $C_{25}$ and $C_{76}$ in black) as their original forms. 
At $t=1$, the CNOT components in $C_{2\_134}$ and $C_{6\_578}$ are executed concurrently on two segments of GHZ states (orange and deep magenta segments in \ref{fig:motivation_with_example}(d), respectively), while at $t=4$, the CNOT components in $C_{4\_1378}$ are executed concurrently on a new GHZ state prepared between $t=1$ and $t=4$ (green segment in \ref{fig:motivation_with_example}(d)).
%
%
In contrast, the other gates are executed individually by bringing their qubits together using SWAP gates  ($t=2$ and $t=3$ in \ref{fig:motivation_with_example}(d)).
%
We omit the details of scheduling and routing here and will discuss them in later sections.

\vspace{0.3em}
\section{Hybrid Computing Paradigm}

Chiplet architecture is a highly scalable architecture that can provide massive qubit resources for NISQ devices. 
This 
not only leads to a leap of computing scale, but can also revolutionize the quantum computing model by making ancillary qubits more affordable.
%
When combined with the mid-circuit measurement technology, these ancillary qubits can be utilized to increase the concurrency of circuit execution, which is particularly crucial when operating at the scale enabled by the chiplet architecture.
This is because as the scale increases, program execution time naturally increases, thereby posing a considerable challenge to program fidelity due to qubit decoherence.
While this can be mitigated by the hardware advancements aimed at extending coherence time, we emphasize the vital role of software optimization, which addresses the problem in a more cost-effective manner.

To take advantage of the ancillary qubits, we extend the circuit model to a hybrid computing paradigm that allows different manners of computation on different types of qubits.
Specifically, computation on the ancillary qubits is performed in a measurement-based manner, while computation on the regular qubits, which we call \emph{data qubits}, remains in the purely gate-based manner.
This hybrid paradigm allows the ancillary qubits to facilitate gate execution between data qubits through a communication protocol depicted in Fig.~\ref{fig:cat_TP}.
This protocol enables simultaneous execution of control gates that share the same control qubit by consuming pre-established GHZ states among the ancillary qubits through 1-qubit measurements.
Given the prevalence of these controlled gates in quantum programs, the integration of additional ancillary qubits can significantly enhance program concurrency.

\begin{figure}[h!]
    \centering
    \includegraphics[width=0.47\textwidth]{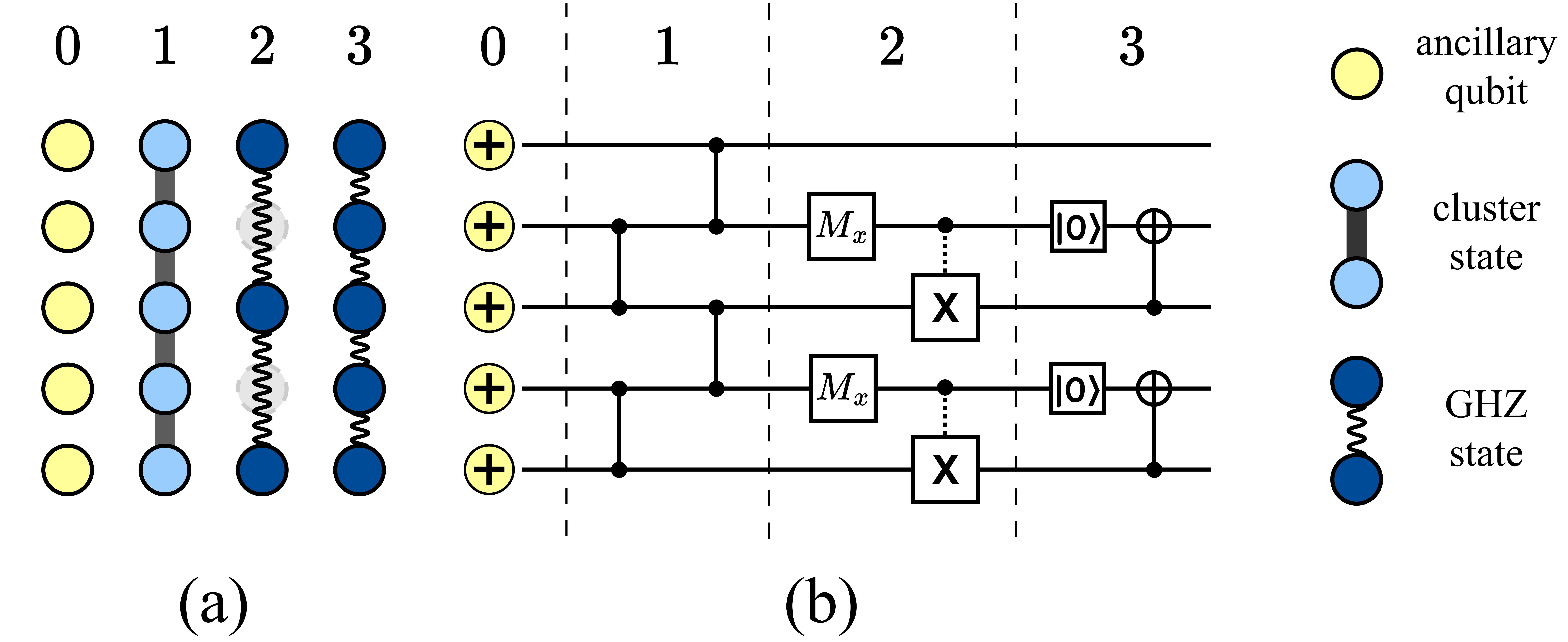}
    \caption{Efficient preparation of GHZ states.}
    \label{fig:cat_preparation}
\end{figure}

An efficient implementation of the protocol in Fig.~\ref{fig:cat_TP} calls for a fast preparation of GHZ states.
This can be realized by manipulating the ancillary qubits in a measurement-based manner.
%
%
Traditionally, a straightforward preparation via chaining the qubits with CNOTs would result in a significant latency which is proportional to the number of entangled ancillary qubits.
To mitigate this issue, we propose an efficient preparation strategy that allows a constant-time preparation by leveraging the mid-circuit measurement technology.
In particular, this is achieved by three steps, as illustrated in Fig.~\ref{fig:cat_preparation}.
First, we prepare an $n-$qubit cluster state in a highly parallel manner.
Second, we perform measurements on half of the entangled qubits, forcing the other half of the qubits to form a $\frac{n}{2}-$GHZ state by applying necessary Pauli corrections.
Third, if necessary, the measured ancillary qubits can be re-entangled by applying parallel CNOT gates controlled by the $\frac{n}{2}-$GHZ state, thus forming a larger GHZ state.

\begin{figure}[h!]
    \centering
    \includegraphics[width=0.47\textwidth]{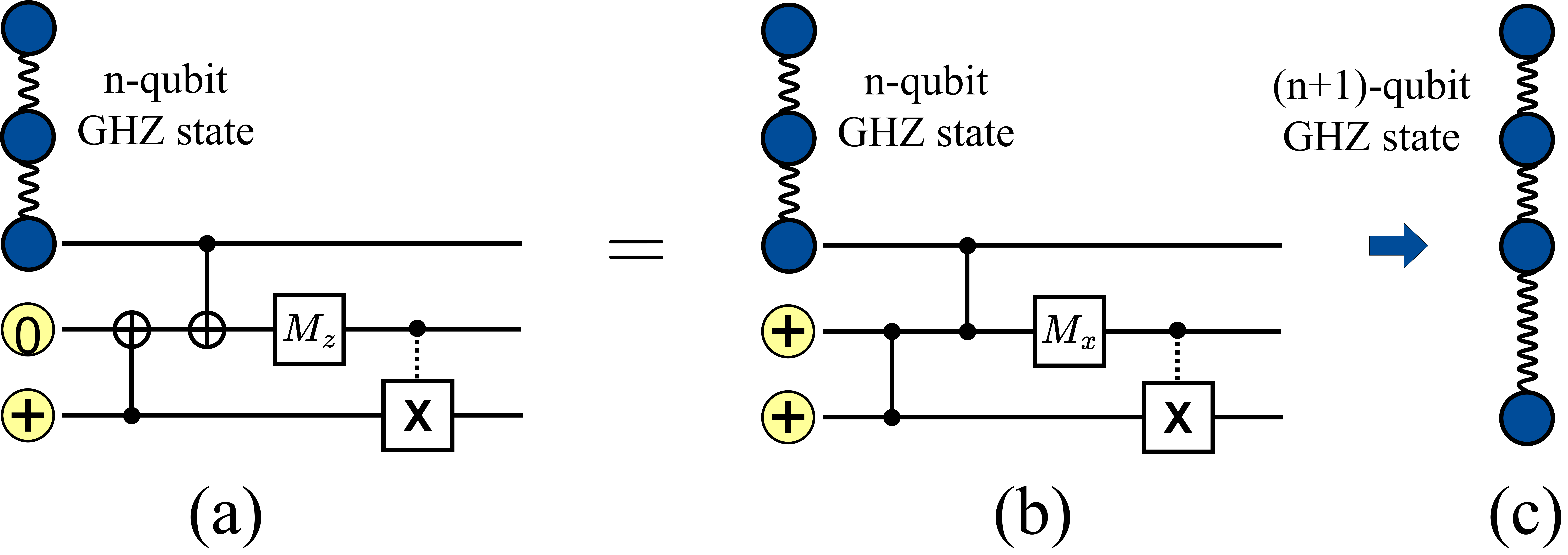}
    \caption{Extension of GHZ states.}
    \label{fig:GHZ_extension}
\end{figure}

The reason behind is that a GHZ state can be extended by the circuit in Fig.~\ref{fig:GHZ_extension}. Specifically, the n-qubit GHZ state $\ket{+}_n \equiv \ket{0}^{\otimes n} + \ket{1}^{\otimes n}$ and the 1-qubit $\ket{+}$ state apply two CNOT gates on another $\ket{0}$ state, leading to a state 
\[\ket{0}^{\otimes n}\ket{0}\ket{0} + \ket{1}^{\otimes n}\ket{0}\ket{1} +  \ket{1}^{\otimes n}\ket{1}\ket{0} +
\ket{0}^{\otimes n}\ket{1}\ket{1}. \]
Then measuring the another qubit results in either  $\ket{0}^{\otimes n}\ket{0}+\ket{1}^{\otimes n}\ket{1}$ with an outcome 0, or $\ket{0}^{\otimes n}\ket{1}+\ket{1}^{\otimes n}\ket{0}$ with an outcome 1. The former state is an (n+1)-GHZ state as shown in Fig.~\ref{fig:GHZ_extension}(c), and the latter one can also be transformed to an (n+1)-GHZ state with a Pauli correction. Since the circuit in Fig.~\ref{fig:GHZ_extension}(a) is equivalent to that in Fig.~\ref{fig:GHZ_extension}(b). This recursively generates the circuit in Fig.~\ref{fig:cat_preparation}, which can be regarded as the generation of a cluster state followed by subsequent measurements and Pauli corrections.
When substituting the 1-qubit $\ket{+}$ state with a multi-qubit GHZ state, this circuit can also be used to merge two GHZ states.

\begin{figure}[h!]
    \centering
    \includegraphics[width=0.45\textwidth]{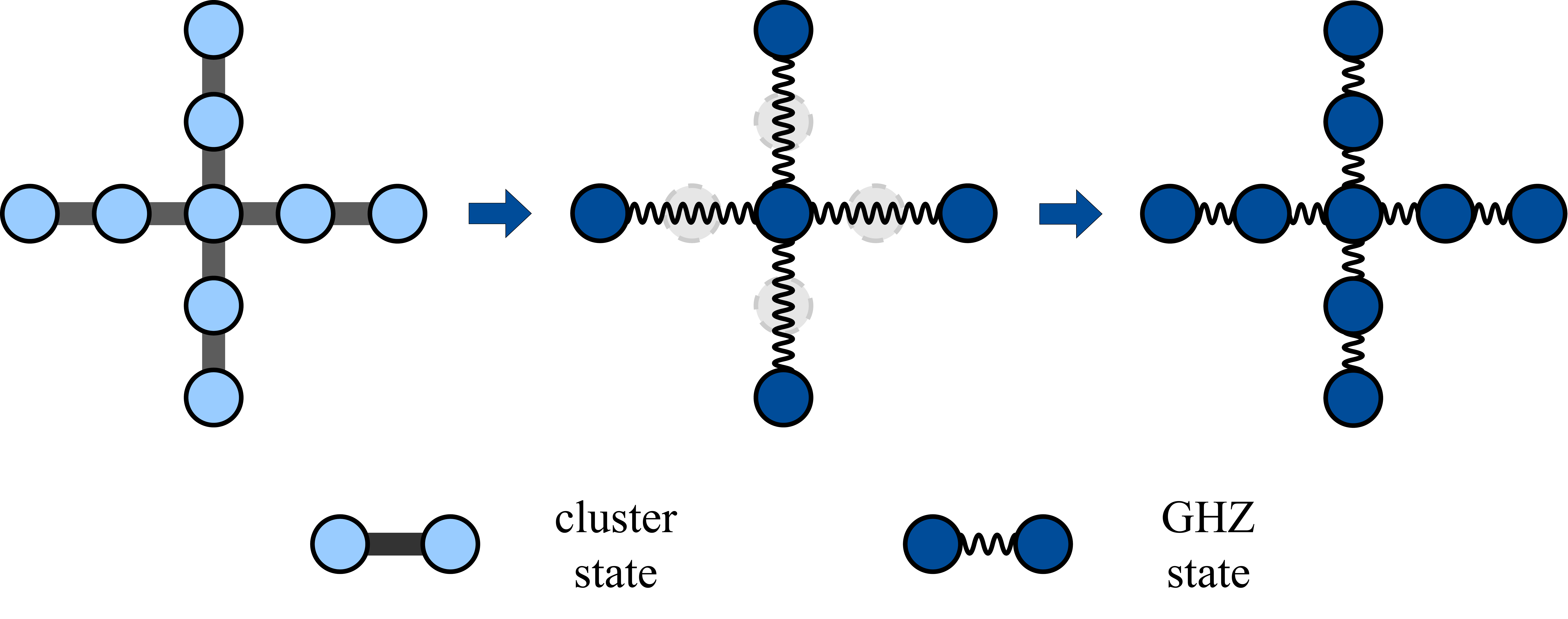}
    \caption{Beyond linear cluster state.}
    \label{fig:nonlinear_cluster}
\end{figure}

This method is applicable for generic qubit layouts, not limited to linear cluster states~\cite{linear_cluster}. For example, a cluster state among the qubits lying on a cross-shaped graph can be transformed to a GHZ state through the process shown in Fig.~\ref{fig:nonlinear_cluster}. 
This lays a foundation for the formation of highway in subsequent sections.

\section{Efficient Highway Mechanism}

When it comes to the allocation of ancillary qubits, there are three challenges we need to address.
First, the implementation of the fast GHZ state preparation is constrained by the hardware connectivity. 
The 2-qubit gates in Fig.~\ref{fig:cat_preparation} can be conducted only between qubits that are adjacent to each other in the coupling graph of chiplets. 
Second, the presence of ancillary qubits inevitably reduces the number of qubits that can be used as data qubits since the total number of qubits is limited.
Hence we need to minimize this qubit overhead so that it does not significantly affects the scale up of computing on the chiplets.
Third, as cross-chip connections have lower fidelity than the on-chip ones, the layout of ancillary qubits should be designed to take this heterogeneity into account and minimize the use of cross-chip connections.

To resolve the first challenge, ancillary qubits should be allocated close to each other on the hardware.
For example, a series of adjacent ancillary qubits can be allocated to form consecutive paths as illustrated in Fig.~\ref{fig:nonlinear_cluster}, so that 2-qubit gates can be directly applied among them.
%
%
We refer to this layout as a \emph{highway} to indicate its speedup for gate execution, and refer to the ancillary qubits forming the highway as \emph{highway qubits}.
The highway should span across chiplets to make sure that it can be accessed easily by all data qubits and facilitate communications between distant qubits.
To maintain the proximity between highway qubits, we allocate the highway qubits prior to the computation and fix their layout throughout the computation.

\begin{figure}[h!]
    \centering
    \includegraphics[width=0.47\textwidth]{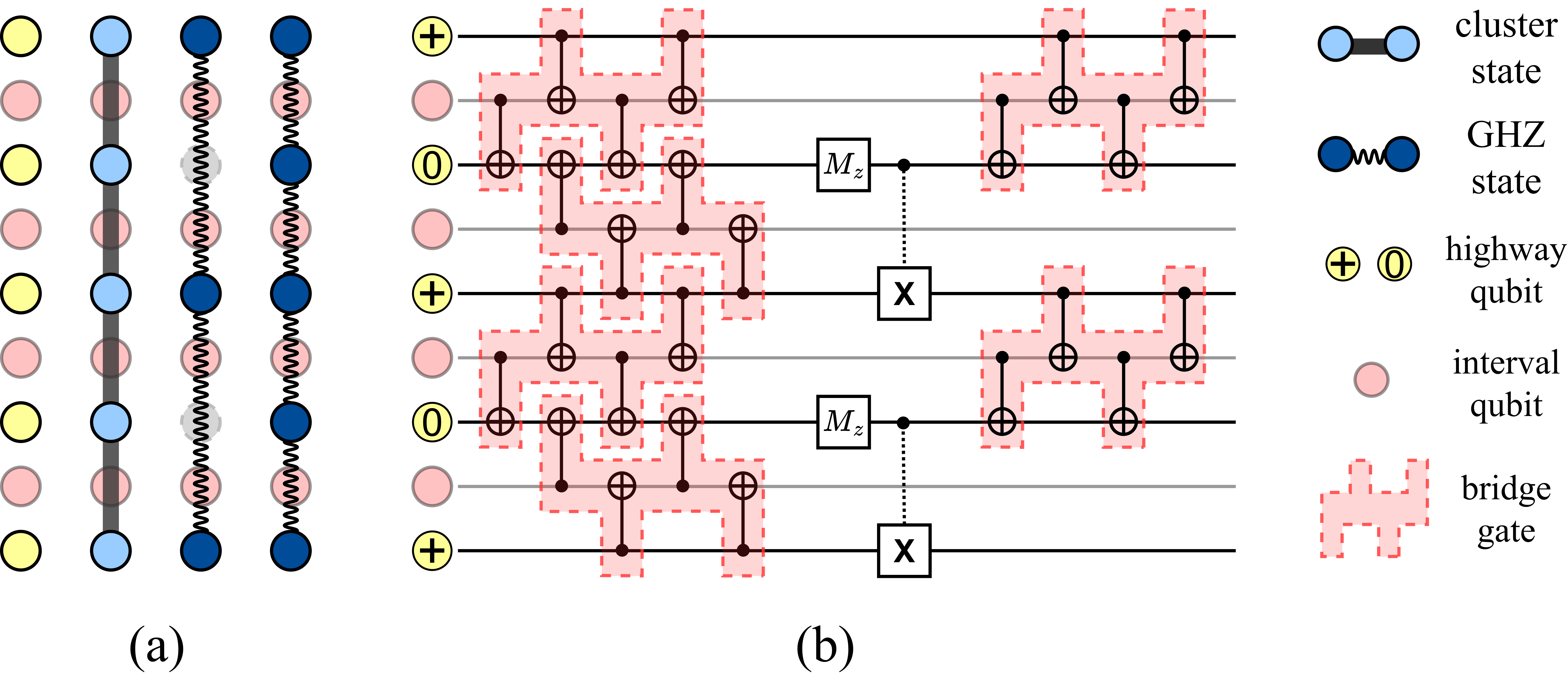}
    \caption{Interleaving highway structure (a) with its efficient GHZ state preparation (b).}
    \label{fig:highway_interleaving}
\end{figure}

The second challenge can be tackled by making the highway structure sparser by interleaving highway qubits with data qubits,
%
as shown in Fig.~\ref{fig:highway_interleaving}(a).
%
The efficient entanglement generation among them can be achieved with bridge gates, with the concrete circuit shown in Fig.~\ref{fig:highway_interleaving}(b).
%
%
The latency of this process is determined by the maximum number of bridge gates each qubit is involved. Consequently, a potential bottleneck arises at the crossroads where different paths intersect.
To mitigate this, we maintain a dense arrangement of highway qubits around the crossroads, only employing an interleaving pattern in the non-crossroad areas.

As for the third challenge, since each cross-chip gate can introduce errors equivalent to several on-chip gates, it is imperative to minimize their usage during the preparation of the highway entanglement.
Similar to that around the crossroads, we can alleviate this concern by maintaining dense qubit arrangements on the edges of each chiplet, 
so that the entanglement among highway qubits on different chiplets can be created by a direct CNOT instead of a bridge gate.
This tradeoff between the qubit overhead and highway performance can be adjusted flexibily based on the performance disparity between on-chip and cross-chip links.

\begin{figure}[h!]
        \centering
        \includegraphics[width=0.47\linewidth]{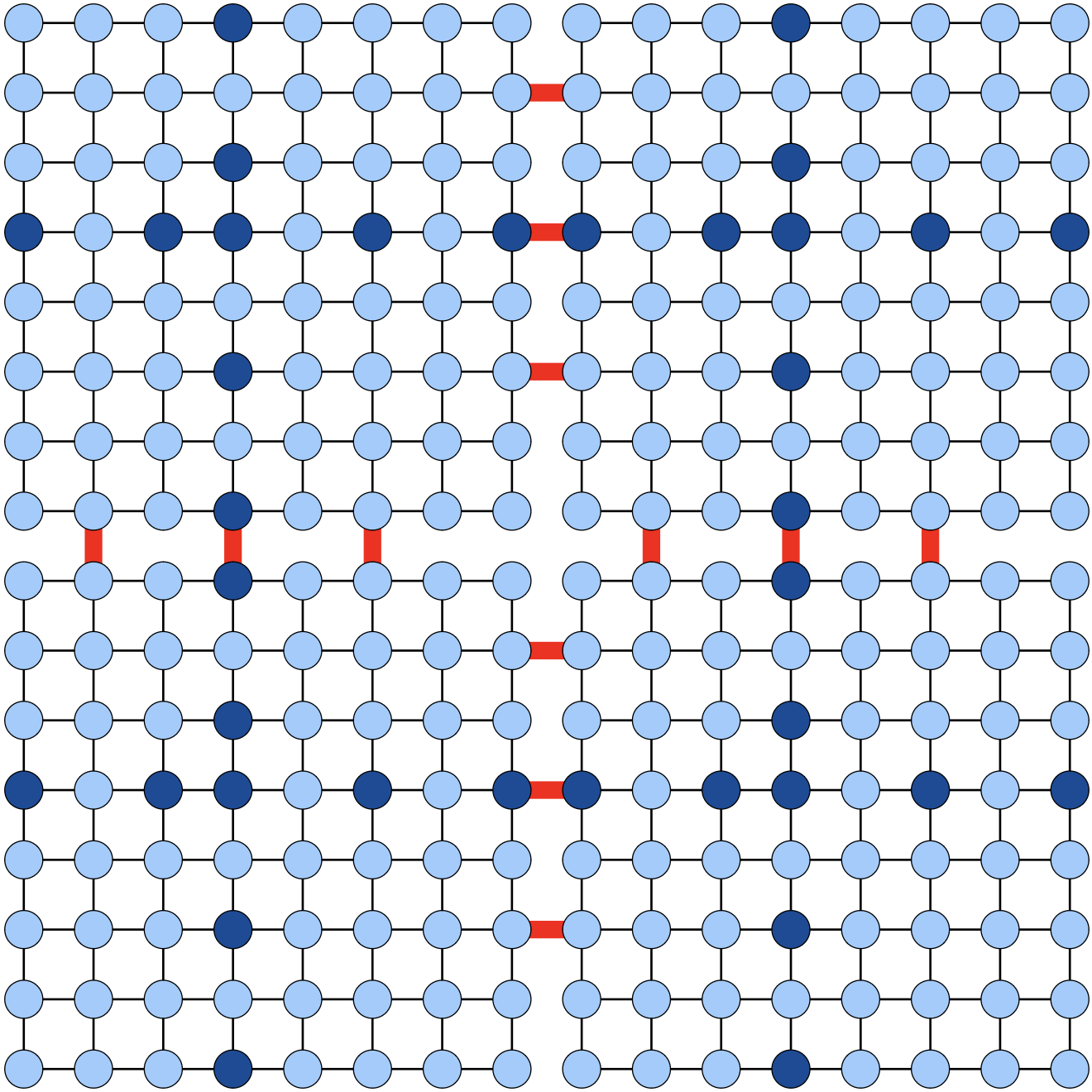} 
        \hspace{10pt}\includegraphics[width=0.47\linewidth]{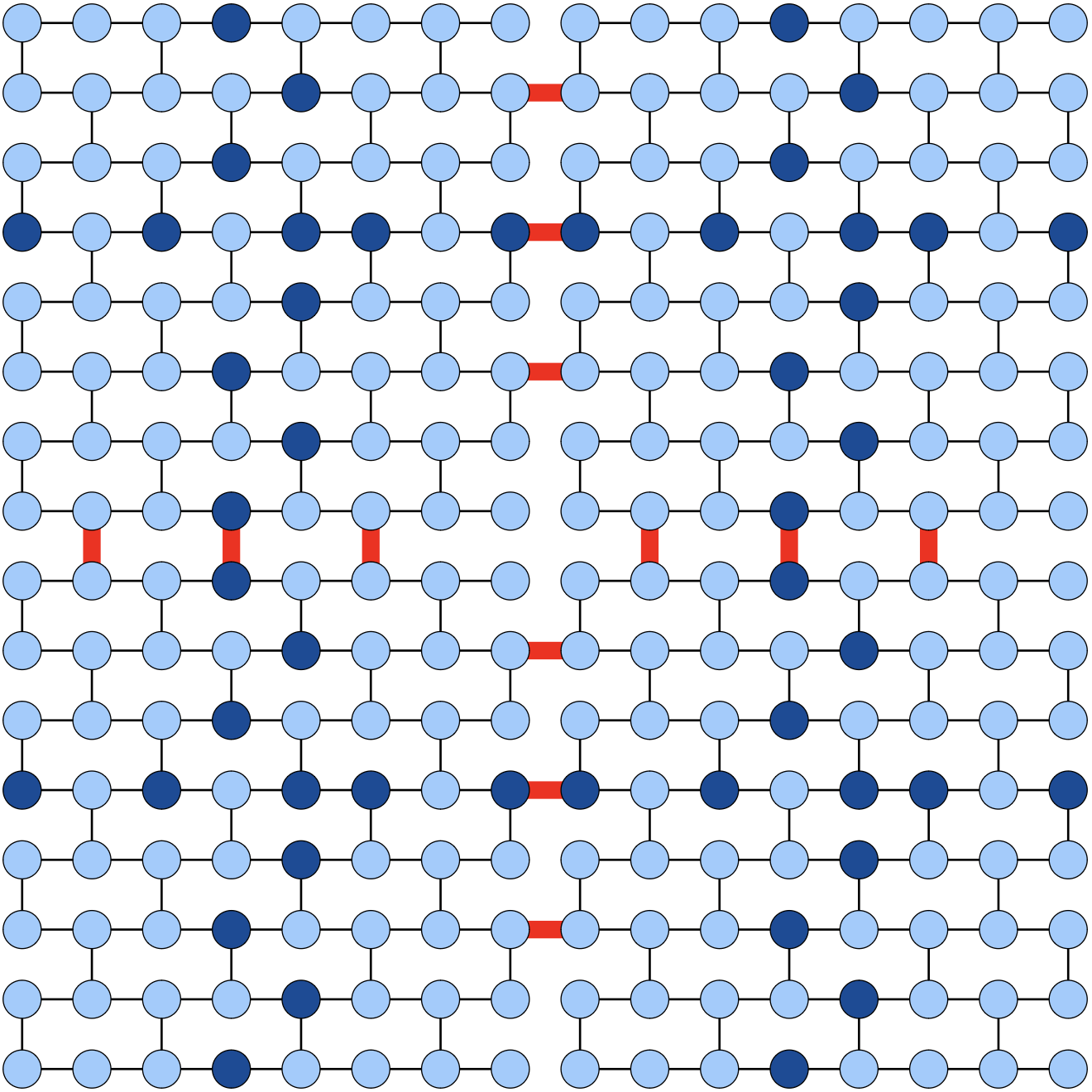}\\
        \hspace{-0pt}(a)\hspace{120pt}(b)
        \caption{Highway configuration on square and hexagon chiplets.}
        
        \label{fig:highway_configuration_example}
\end{figure}

In compliance with the conditions set forth by the insights above, we can construct the highway in a roughly mesh-like configuration, with the selection of horizontal and vertical paths ensuring that they encounter the cross-chip links at the boundaries. As an example, Fig.~\ref{fig:highway_configuration_example}(a)(b) illustrate the highway configurations when the chiplets are in a square and hexagon structure. In the example, the highways on each chiplet have the same configuration, since the periodicity of the mesh aligns with the the width of each chiplet. However, this is not mandatory. In general cases, we can consider connections within and beween all chiplets collectively, allowing the mesh's periodicity to be flexibly determined based on the affordability of ancillary qubtis.

The GHZ state on the highway is prepared by identifying a set of critical highway qubits, i.e., those at crossroads and those at the ends of a path, then entangling them in the way of  Fig.~\ref{fig:highway_interleaving}. 
For each pair of critical qubits, we first count the number of highway qubits between them. If the number is odd, we connect them with the GHZ preparation circuit (Fig.~\ref{fig:highway_interleaving}) with either bridge gates or direct CNOT gates. If the number is even, we turn it into the odd case by first conducting a CNOT (preferred) or bridge gate between one of the critical and the highway qubit next to it. After that, we measure the necessary qubits and do the Pauli corrections. Finally, for the measured qubits, if they serve as a highway entrance (will be explained in the next section), they will be re-entangled by the nearest unmeasured highway qubits which so far has the least number of gates during the GHZ state preparation process.

\section{Routing and Scheduling Strategies}

To optimize the utilization of highway resources, it is crucial to reduce the communication latency with an efficient implementation of the \emph{highway gates}, including an efficient routing and scheduling.
By highway gates, we refer to the gates executed  via the highway protocol. Since each highway gate consists of multiple 2-qubit controlled gates, we call each of these 2-qubit controlled gates a \emph{gate component} of the highway gate. 

\subsection{Routing}
There are two major challenges for routing, one arising from the hardware constraint and the other from the highway mechanism.
First, due to the connectivity constraints, the execution of a highway gate requires each of its control and target qubits to be adjacent to a highway qubit. We call that highway qubit a \emph{highway entrance} for that control or target.
Second, the efficient preparation of each GHZ state involves CNOTs along a path of qubits, and the paths of GHZ states required by different highway gates should not overlap.
We refer to these paths occupied by corresponding highway gates as their \emph{highway paths}.
For example, in Figure~\ref{fig:motivation_with_example}, 
the highway paths of $C_{2\_134}$ (orange gate in \ref{fig:motivation_with_example}(b)) and $C_{6\_578}$ (deep magenta gate in \ref{fig:motivation_with_example}(b)) are disjoint to each other.

The first challenge necessitates an efficient \emph{local routing} that brings qubtis of highway gates towards the highway. Our strategy is that each qubit involved in the highway gates is routed to a nearby highway entrance that enables its earliest execution, which is implemented by inserting SWAP gates along a path that connects the qubit to the highway entrance. To find an optimal highway entrance, each data qubit in the highway gates searches for candidate entrances from its nearby highway qubits. Given the current accumulated depth on the data qubit, its arrival time to each candidate entrance can be obtained by finding the shortest paths to the candidate, denoted as $t_{arr}$. Compared with the next available time of each candidate, denoted as $t_{ava}$, we can get the data qubit's earliest execution time on each candidate entrance, which is $t_{exe} =\max\{t_{arr}, t_{ava}\}$. Then the candidate with the earliest $t_{exe}$ is selected as the entrance for the data qubit.
This entrance assignment is performed on data qubits in an ascending order of their shortest distance to the highway to minimize the conflicts between the local routing paths.

The second challenge can be addressed by an efficient \emph{highway routing} that finds the optimal highway path for each highway gate.
Our insight is that the length of the highway path for a highway gate can be minimized by maximizing the reuse of highway paths among its gate components. Specifically, each component of the highway gate is assigned a highway path that occupies the least number of additional highway qubtis. This can be achieved by a shortest path search between the component's control entrance and target entrance, with the edge weights along the paths occupied by other components set as 0.
In this process, if 
the highway path of a component has to occupy a highway qubit that is already occupied by components of a different highway gate, then the component is temporarily not executable and has to wait for those occupied highway qubits to be released.
This highway path assignment is performed for all components of each highway gate, in descending order of the number of components in the highway gates, resulting in a set of disjoint highway paths occupied by different highway gates.

\subsection{Scheduling}

During the highway communication, GHZ states are periodically prepared and consumed, with each round referred to as a \emph{highway shuttle}. To reduce the latency of highway protocols, we allow a dynamic period for each shuttle based on the traffic demand so that more gate components can share the highway in the same shuttle. Each component of the highway gates is scheduled to the earliest possible highway shuttle given the available highway paths and entrances of each shuttle. The period of each shuttle is determined by waiting on more gates until a new shuttle is needed by gates requiring conflict highway paths. After a new shuttle is scheduled, subsequent highway gates can still be added to this shuttle if they arrive at the highway within the shuttle period, but they can no longer increase the period.

\begin{figure}[h]
    \centering 
    \includegraphics[width=0.47\textwidth]{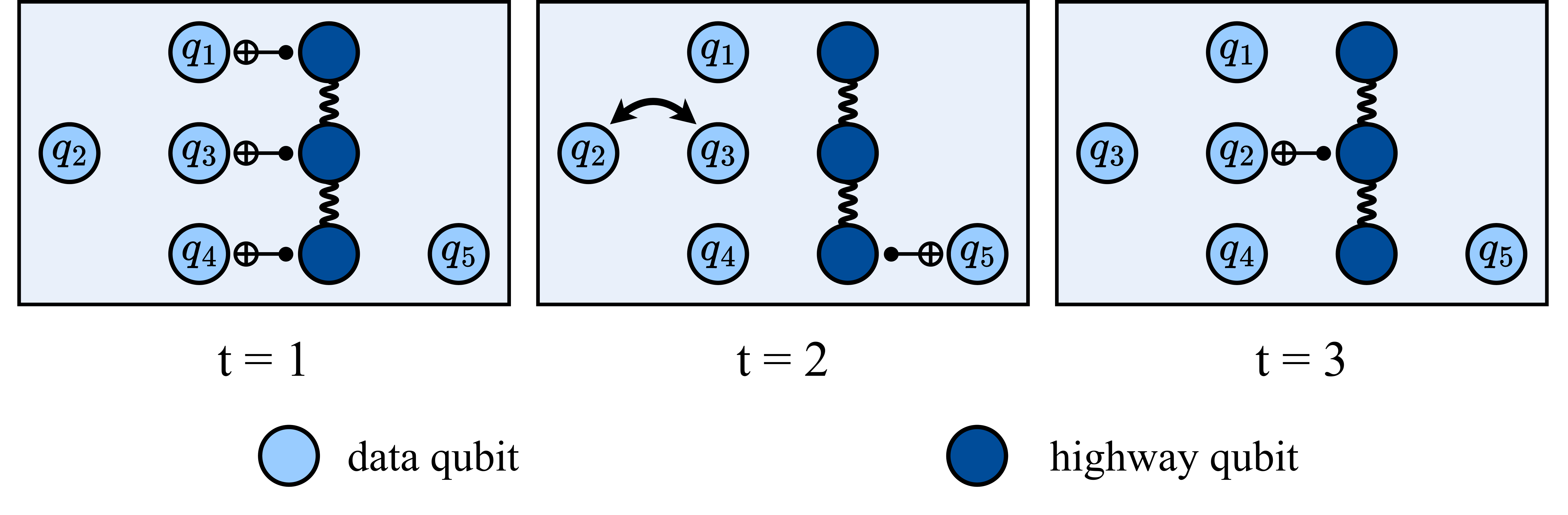}
    \setlength{\abovecaptionskip}{0.cm}
    \caption{Dynamic period of highway communication according to traffic demand.}
    
    \label{fig:dynamic_period}
\end{figure}

Figure~\ref{fig:dynamic_period} illustrates the dynamic period with an example of 5 data qubits contending for 3 highway entrances. 
Although only 3 of the qubits can be executed at $t=1$, this shuttle waits until $t=3$ so that the other 2 qubits can be executed at $t=2$ and $t=3$. Then this shuttle will be ended by destroying the entanglement with measurements on the highway qubits, and the highway qubits become free to prepare for the next shuttle. 
This dynamic period allows different target qubits of a highway gate to share the same highway entrance at different times, thus reducing the total number of shuttles required. 

To maximize the number of gate components in the highway gates, we allow rewriting of the circuit to aggregate the controlled gates sharing the same control qubit. The aggregated multi-target gates among all executable gates are ranked by their numbers of component. Those with the most gate components will be scheduled as highway gates, while the others are scheduled as regular 2-qubit gates off the highway.

\section{Evaluation}\label{sect:eval}


\subsection{Experiment Setup}\label{sect:expset}

\paragraph{Baseline} 
%
We implement the baseline by applying state-of-the-art compiler in Qiskit~\cite{Qiskit} with the highest optimization level 3. 
The coupling graph passed to the Qiskit compiler contains both the on-chip and cross-chip links.
We execute the same circuits on the same chiplet settings (chiplet size, chiplet array size, coupling structure, etc.) for both the baseline and our framework, with the circuit sizes determined by the numbers of data qubits in our framework.

\paragraph{Benchmark Programs} We select quantum Fourier transform (QFT), quantum approximate optimization algorithm (QAOA), variational quantum eigensolver (VQE) and Bernstein Vazirani (BV) algorithm as our benchmarks.
For QAOA,
we choose the graph maxcut problem on randomly generated graphs. Specifically, the graphs are generated by randomly connecting half of all its possible edges.
For VQE, we follow the commonly used full-entanglement ansatz, which proves to be an expressive ansatz \cite{qiskit_vqe, vqe_ansatz}.
For BV, we select the secret strings randomly, with approximately half of the digits being 0 and half being 1. In table~\ref{tab:benchmark}, we list the number of data qubits, the number of total qubits, the size of each chiplet and the size of the chiplet array for these benchmarks in subsequent experiments including those in sensitivity analysis. 

\begin{table}[h]
    \centering
      \caption{Architecture Settings.}
      
    \resizebox{0.47\textwidth}{!}{
        \renewcommand*{\arraystretch}{1.2}
        \begin{tabular}{|p{3.2cm}|p{2.2cm}|p{0.9cm}|p{0.9cm}|p{0.9cm}|}  \hline
        name-\#data qubits  & coupling  & \#total  & chiplet & chiplet   \\
         & structure &  qubits &  size & array  \\ \hline

        program-261 & square & 324 & 6x6 & 3x3\\ \hline
        program-360 &  square & 441 & 7x7 & 3x3\\
          \hline
        program-495 &  square & 576 & 8x8 & 3x3\\ \hline
        program-630 &  square & 729 & 9x9 & 3x3\\  \hline

        program-160 &  square & 196 & 7x7 & 2x2\\  \hline
        program-240 &  square & 294 & 7x7 & 2x3\\  \hline
        program-480 &  square & 588 & 7x7 & 3x4\\  \hline

        program-420/366/288 &  square & 486 & 9x9 & 2x3\\  \hline

        program-312 &  hexagon & 384 & 8x8 & 2x3\\  \hline
        program-351 &  heavy square & 432 & 8x8 & 3x3\\  \hline
        program-336 &  heavy hexagon & 480 & 8x8 & 3x4\\  \hline
         
        \end{tabular}
    }
    \label{tab:benchmark}
\end{table}

\paragraph{Metric} The first metric we consider is the depth of the compiled circuit, as it represents the latency of program execution and is essential for the mitigation of demand for qubit decoherence time. 
%
%
When couting the circuit depth, we only focus on 2-qubit CNOT gates and measurements, while ignoring the 1-qubit gates and classical operations as they are much faster. 
We count each measurement operation as multiple depths to manifest its longer latency than gate operations.
%
The second metric we consider is the number of the most error-prone operations, including on-chip CNOT gates, cross-chip CNOT gates and measurements.
%
The overall error rate is as below when these error rates are small
\[
\mathrm{error\_rate} = \enspace 1 - \prod_i (1-p_i)^{n_i}
=\enspace \sum_i n_i p_i + O(p_i^2)
\]
where $n_i$ is the number of the $\mathrm{i^{th}}$ operation, and $p_i$ is its error rate. Hence we address the disparity in the error rates of different operations by defining an \emph{effective number of CNOT gates} as below, abbreviated as `\#eff\_CNOT', 
\begin{align*}
\# \mathrm{eff\_CNOTs} = &\enspace \# \mathrm{on\_chip\_CNOTs} \\
&+ \frac{p_{cross}}{p_{on}}\times\# \mathrm{cross\_chip\_CNOTs} \\
&+ \frac{p_{meas}}{p_{on}}\times\# \mathrm{measurements} 
\end{align*}
where $p_{on}$, $p_{cross}$ and $p_{meas}$ are the error rates of on-chip CNOTs, cross-chip CNOTs and measurements.

\begin{figure}[h!]
    \centering
    \includegraphics[width=0.41\textwidth]{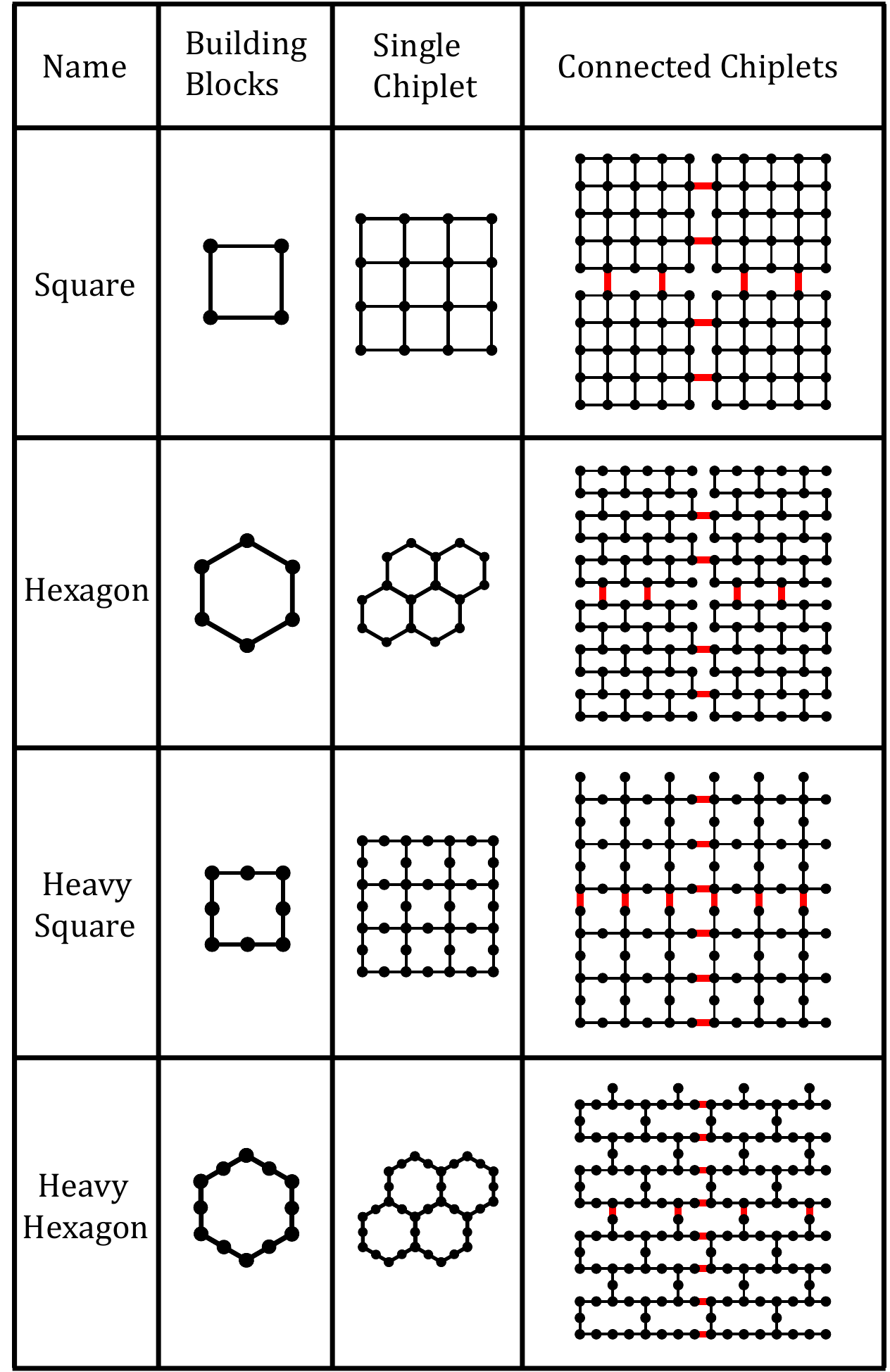}
    \caption{Different structures for chiplet architecture. Black lines indicate on-chip connections, red lines indicate cross-chip connections.}
    \label{fig:Couple}
\end{figure}

\paragraph{Coupling Structure} Our experiments are performed for chiplets of various structures, including square, hexagon, heavy-square and heavy-hexagon. The concrete coupling structures of them are shown in Fig.~\ref{fig:Couple}, with the black edges indicating  on-chip connections and the red edges indicating cross-chip connections.

\begin{table*}[tp]
    \centering
    \caption{The results of \frameworkname\ and its relative performance to the baseline. }
    \resizebox{\textwidth}{!}{
        \renewcommand*{\arraystretch}{1}
        \begin{small}
        \begin{tabular}
        {|p{1.8cm}|p{1.6cm}|p{1.6cm}|p{2.1cm}|p{1.6cm}|p{1.6cm}|p{2.1cm}|p{1.4cm}|} \hline
        {Program} & {Baseline} & {Framework} & {Depth} & {Baseline} & {Framework} & {eff\_CNOTs} & {Highway}    \\ 
        {} & { Depth} & { Depth} & {Improvement} & { eff\_CNOTs} & { eff\_CNOTs} & {Improvement} & { Qubit \%}    \\ 
        \hline
        QFT-261     &   19,282  &  7,504  &  61.1\%  &  325,236  &  216,771  &  33.3\%  &   19.4\%    \\
        \hline
        QAOA-261    &   14,837  &  6,586  &  55.6\%  &  201,637  &  151,120  &  25.1\% &   19.4\%    \\
        \hline
        VQE-261     &   15,725  &  6,784  &  56.9\%  &  261,286  &  180,044  &  31.1\% &   19.4\%    \\
        \hline
        BV-261      &   418  &  31  &  92.6\%  &  1,179  &  960  &  18.6\% &   19.4\%    \\
        \hline

        QFT-360     &   32,086  &  11,189  &  65.1\%  &  582,500  &  451,553  &  22.5\% &   18.4\%    \\
        \hline
        QAOA-360    &   22,757  &  9,735  &  57.2\%  &  389,773  &  300,847  &  22.8\% &   18.4\%    \\
        \hline
        VQE-360     &   26,277  &  10,181  &  61.3\%  &  471,148  &  385,647  &  18.1\% &   18.4\%    \\
        \hline
        BV-360      &   597  &  34  &  94.3\%  &  1,711  &  1,415  &  17.3\% &   18.4\%    \\
        \hline

        QFT-495     &   57,143  &  18,028  &  68.5\%  &  1,048,824  &  827,653  &  21.1\% &   14.1\%    \\
        \hline
        QAOA-495    &   43,478  &  14,175  &  67.4\%  &  716,324  &  507,897  &  29.1\% &   14.1\%    \\
        \hline
        VQE-495     &   47,193  &  16,512  &  65.0\%  &  854,935  &  690,826  &  19.2\% &   14.1\%    \\
        \hline
        BV-495      &   823  &  37  &  95.5\%  &  2,297  &  1,784  &  22.3\% &   14.1\%    \\
        \hline
        
        QFT-630     &   90,535  &  24,138  &  73.3\%  &  1,673,337  &  1,511,568  &  9.7\% &   13.6\%    \\
        \hline
        QAOA-630    &   66,342  &  19,115  &  71.2\%  &  1,171,597  &  914,800  &  21.9\% &   13.6\%    \\
        \hline
        VQE-630     &   75,178  &  21,687  &  71.2\%  &  1,370,750  &  1,296,846  &  5.4\% &   13.6\%    \\
        \hline
        BV-630      &   1,063  &  40  &  96.2\%  &  2,772  &  2,612  &  5.8\% &   13.6\%    \\  
        \hline
        \end{tabular}
        \end{small}
    }
    \label{tab:evaluation}
\end{table*}

\subsection{Experiment Result}
In this subsection, we demonstrate the performance and scalability of our framework. Each measurement is counted as a depth of 2 based on the calibration data of IBM \cite{ibm_experience}. We adopt a rate 7.4 for $p_{cross}/p_{on}$ according to the on-chip fidelity reported recently by IBM \cite{99.77} and the cross-chip fidelity of flip-chip bonds \cite{short_range_hardware_5}. We obtain a ratio $p_{meas}/p_{on}=2.2$ using the same on-chip fidelity and the measurement fidelity reported recently \cite{99.5}. Since these parameters vary across different platforms, we will vary them in the sensitivity analysis later. Moreover, this subsection focuses on the square coupling structure with dense cross-chip links. Different coupling structures and cross-chip sparsity levels will be explored in sensitivity analysis too.

\paragraph{Performance} Table~\ref{tab:evaluation} shows the results of the baseline and \frameworkname\ on a 3x3 array of chiplets, with the size of each chiplet varying from 6x6 to 9x9. It can be seen that \frameworkname\ significantly reduces both the circuit depth and the effective number of CNOT gates compared to the baseline. On average (geomean), the circuit depth is reduced by 70.8\%, and the effective number of CNOTs is reduced by 18.1\%.
Furthermore, as the chiplet size increases, the qubit overhead decreases from 19.4\% to 13.6\%, as can be obtained from Table~\ref{tab:benchmark}.
The outperformance in circuit depth increases with the chiplet size, because the program benefits more from the increased concurrency as the circuit grows larger. The outperformance in the number of effective CNOT decreases with the chiplet size, because the lower percentage of highway qubits result in more demand for local routing of data qubits to access the highway.

\begin{figure}[h!]
        \centering
        \hspace{-15pt}\includegraphics[width=0.86\linewidth]{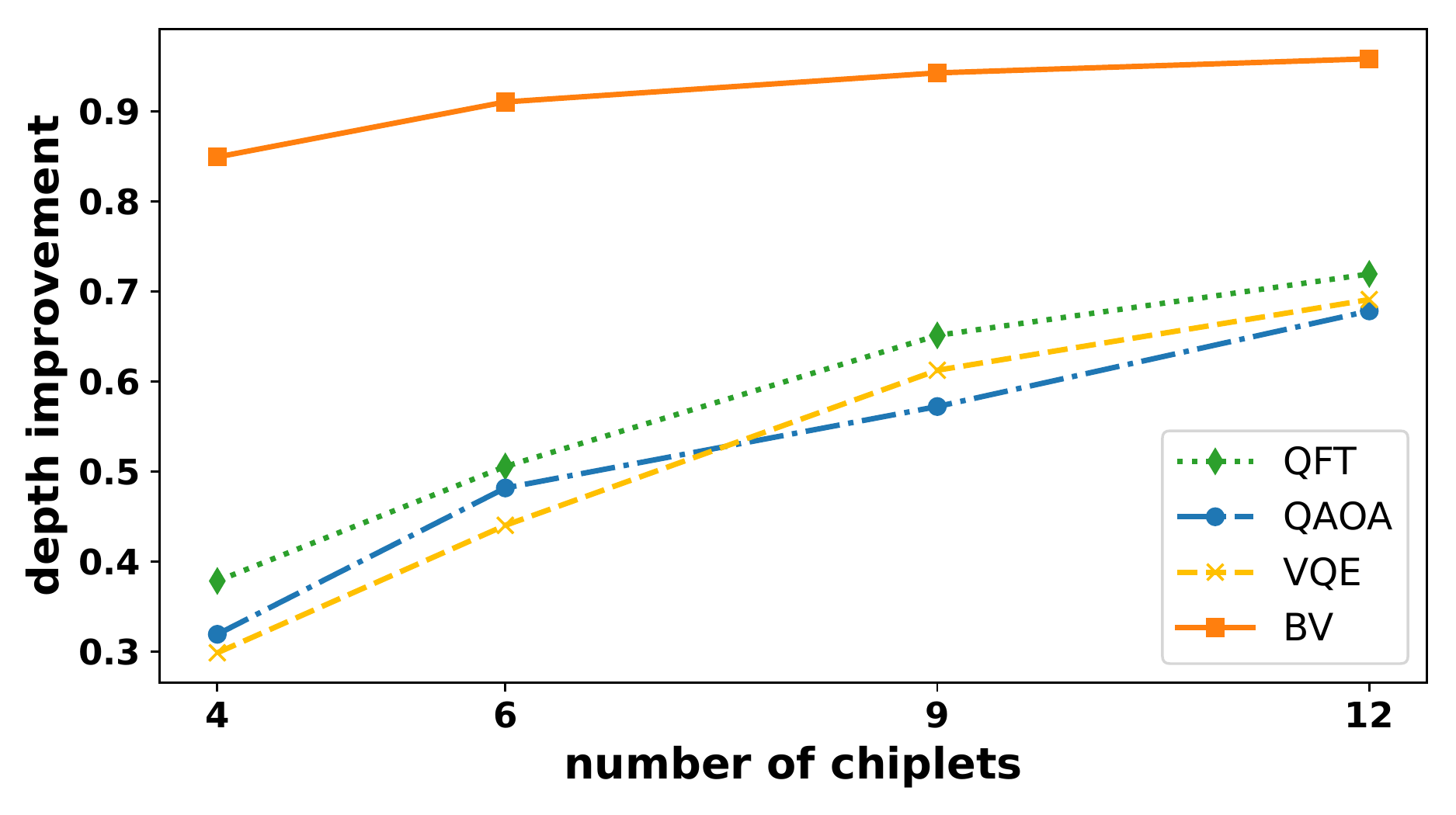} \\
        \vspace{-5pt}
        (a)\\
        \vspace{5pt}
        \hspace{-15pt}\includegraphics[width=0.86\linewidth]{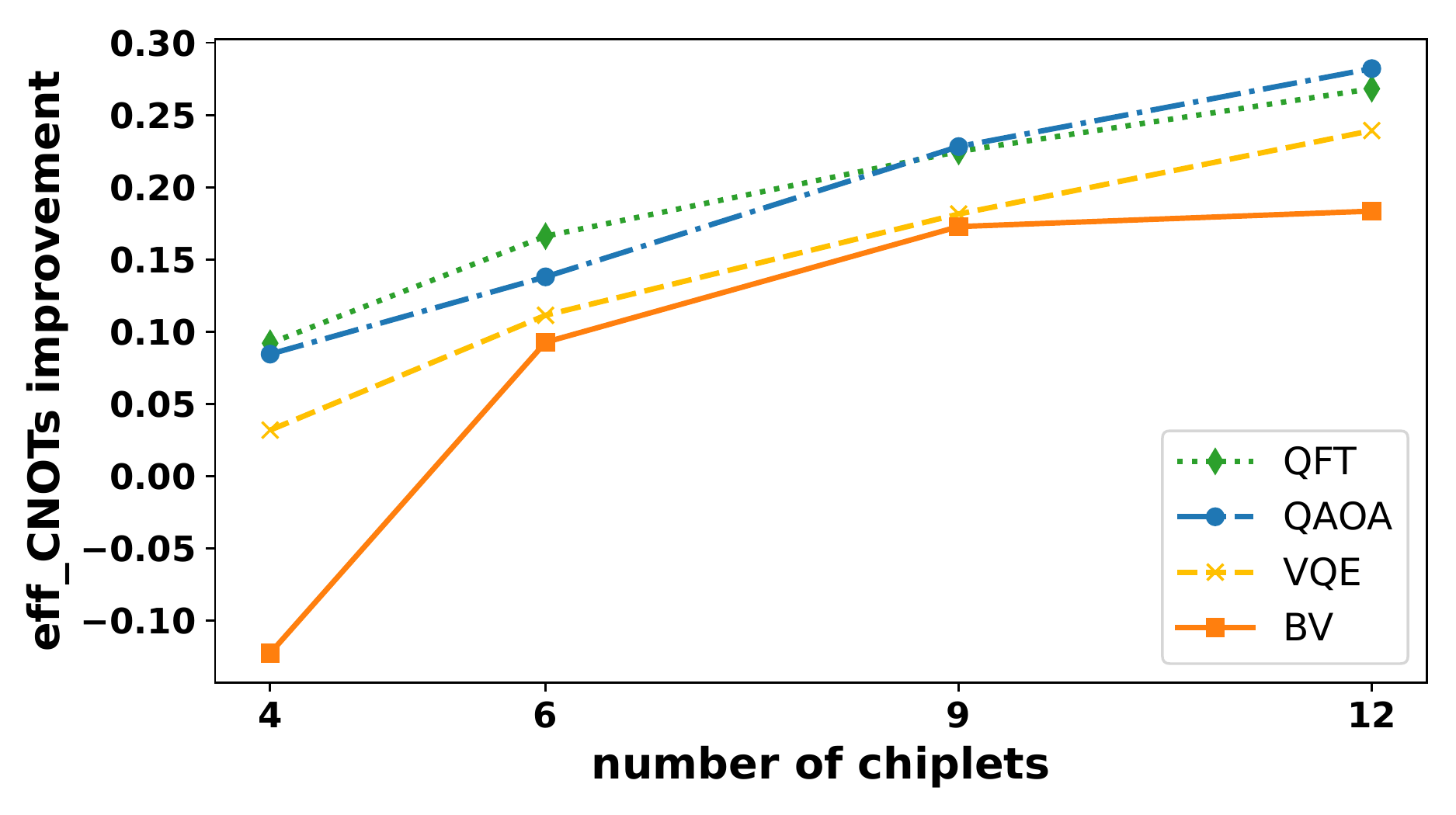}\\
        \vspace{-5pt}(b)
    
        \caption{Improvement in performance for increasing number of chiplets.}
        
        \label{fig:different_chiplet_sizes}
\end{figure}

\paragraph{Scalability}
Besides the demonstration on varying sizes of chiplets, we demonstrate the scalability of our framework by evaluating on varying numbers of chiplets. Fig.~\ref{fig:different_chiplet_sizes} shows the improvements in the circuit depth (\ref{fig:different_chiplet_sizes}(a)) and the effective number of CNOTs (\ref{fig:different_chiplet_sizes}(b)) by \frameworkname\ on 2x2, 2x3, 3x3 and 3x4 chiplet arrays, with the size of each chiplet fixed as 7x7. It can be seen that our framework not only achieves significant improvement in performance, but the improvement also increases as the number of chiplets increases.
This trend suggests that the highway model can potentially play an important role in the way of achieving tens of thousands of qubits with the chiplet architecture.

\subsection{Sensitivity Analysis}
In this section, we further analyze the effects of the various parameters and configurations, including the latency of measurements, the fidelities of measurements and cross-chip gates, the sparsity of cross-chip links, the highway qubit percentage and the coupling structure of chiplets. 

\begin{figure}[h!]
        \centering
        \hspace{-15pt}\includegraphics[width=0.86\linewidth]{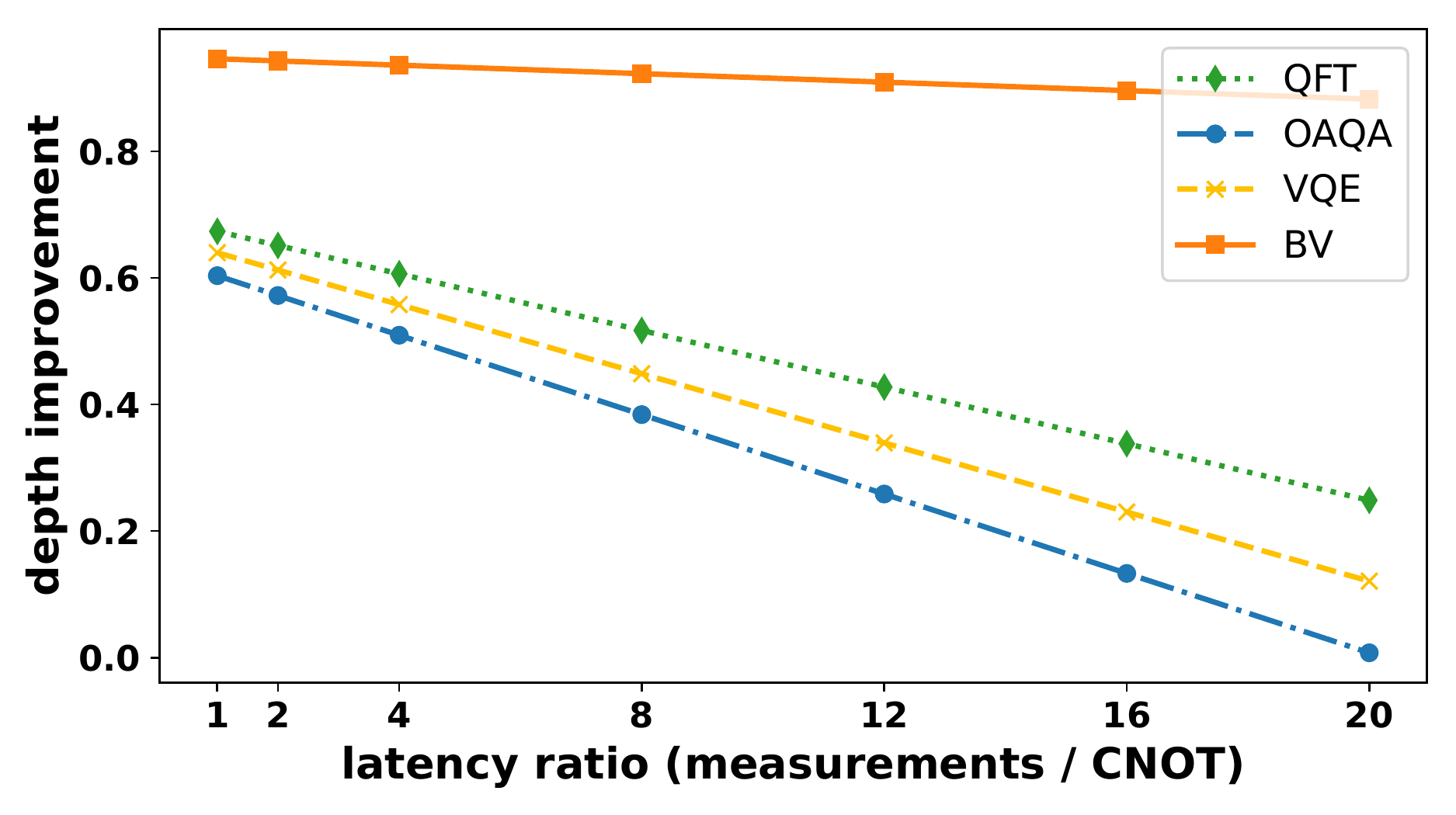} \\
        \vspace{-5pt}(a)\\
        \vspace{5pt}
        \hspace{-15pt}\includegraphics[width=0.86\linewidth]{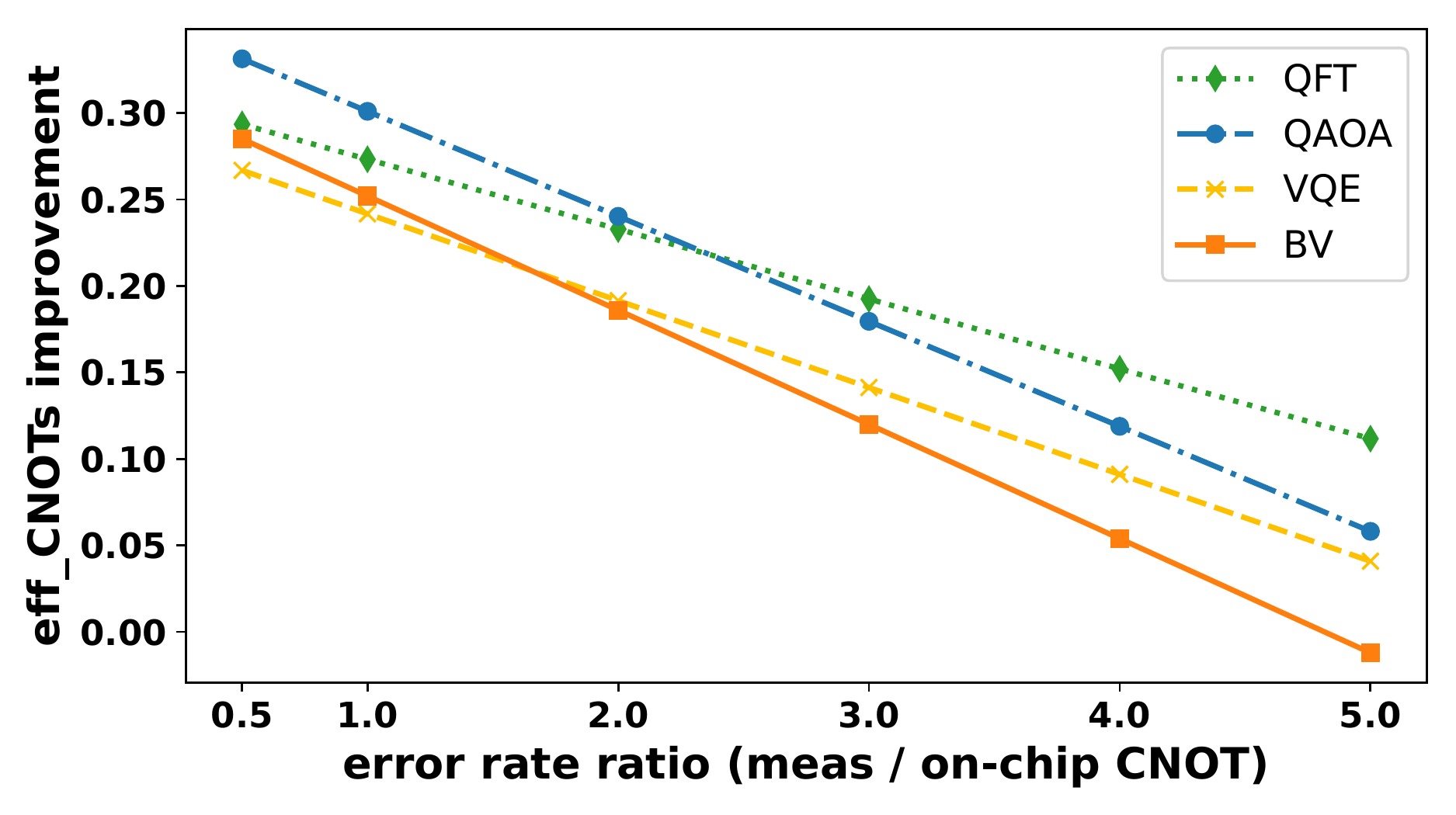}\\
        \vspace{-5pt}(b)\\
        \vspace{5pt}
        \hspace{-15pt}\includegraphics[width=0.86\linewidth]{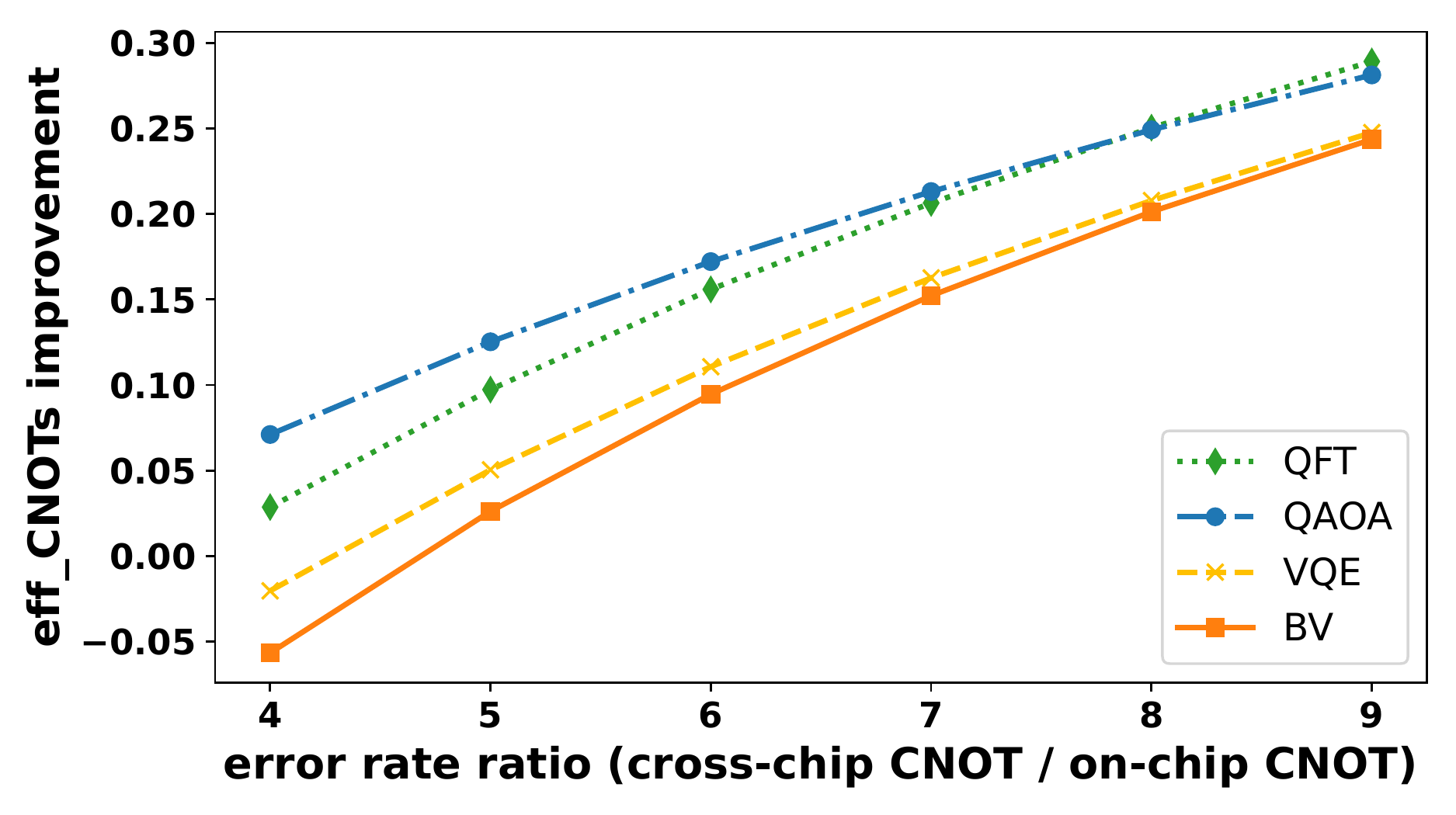}\\
        \vspace{-5pt}(c)
    
        \caption{Improvement in performance for varying latencies and fidelities of measurements and cross-chip CNOTs.}
        
        \label{fig:meas_sensitivity}
\end{figure}

\paragraph{Measurement Latency} We illustrate the effect of measurement latency with a 3x3 array of 7x7 chiplets. 
As can be seen from Fig.~\ref{fig:meas_sensitivity}(a), the improvement in circuit depth exhibits a linear decrease as the measurement latency increases, remaining positive up to a depth of 20. This implies that the outperformance of our framework in circuit depth is not sensitive to the measurement latency. The linearity is because the measurement latency affects the overall circuit depth only by introducing a constant overhead to each highway protocol, as measurements are conducted highly concurrently.

\paragraph{Operation Fidelity} To evaluate the effect of fidelity disparity among different operations, we vary the fidelity ratios of measurements and cross-chip CNOTs to on-chip CNOTs for a 3x3 array of 7x7 chiplets. Fig.~\ref{fig:meas_sensitivity}(b)(c) shows the impact of measurement fidelity and the impact of cross-chip CNOT fidelity on the number of effective CNOT gates, respectively.
The improvement of effective CNOTs decreases with an increased error rate of measurements, remaining positive up to a ratio of 5, while it increases with an increased error rate of cross-chip CNOTs, being positive for ratios larger than 4. These indicate that our framework can outperform the baseline within a wide range of fidelities of different operations.

\paragraph{Cross-chip Sparsity}
\begin{figure}[h!]
        \centering
        \hspace{-15pt}\includegraphics[width=0.86\linewidth]{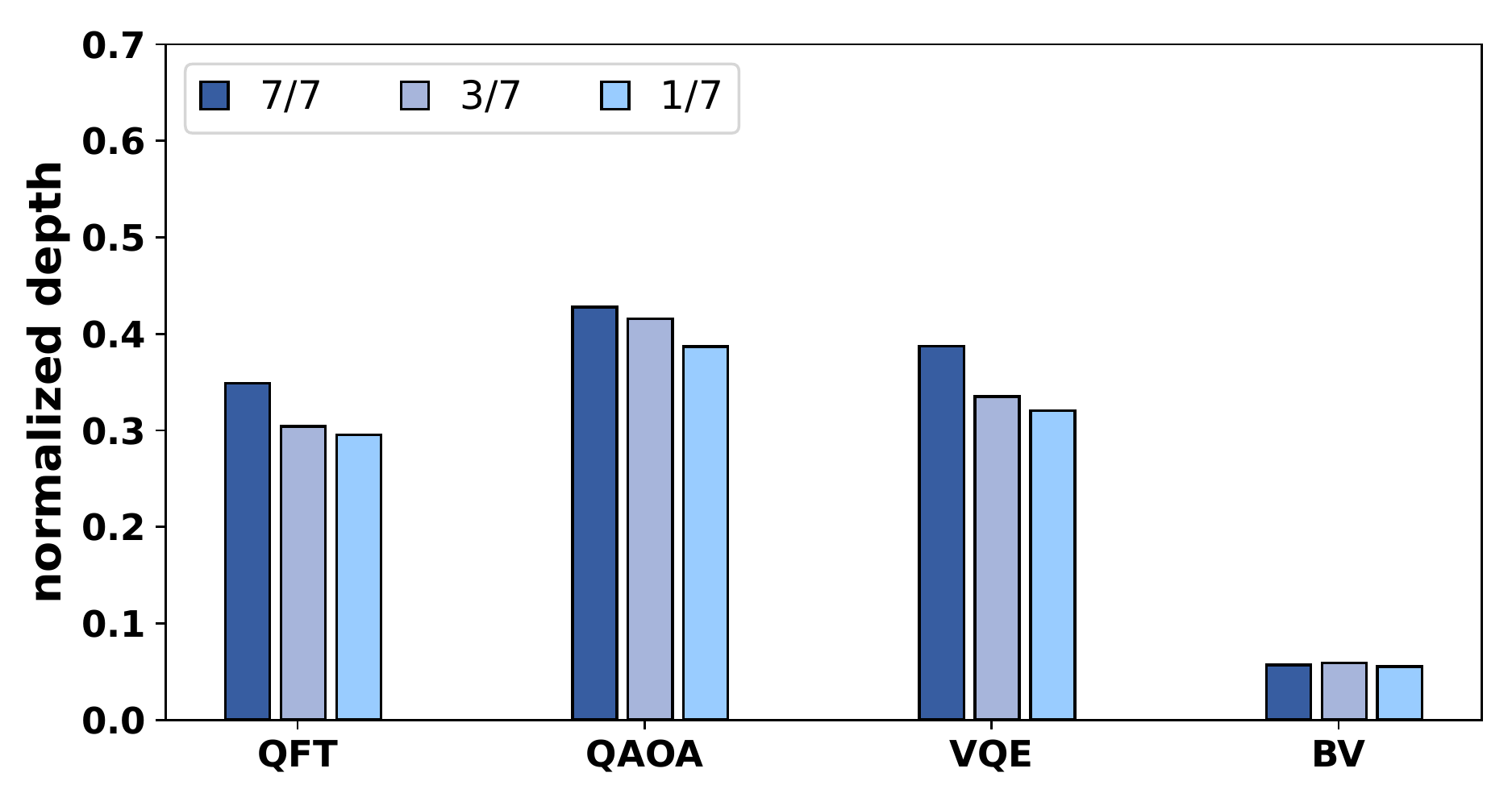} \\
        \vspace{-5pt}(a)\\
        \hspace{-15pt}\includegraphics[width=0.86\linewidth]{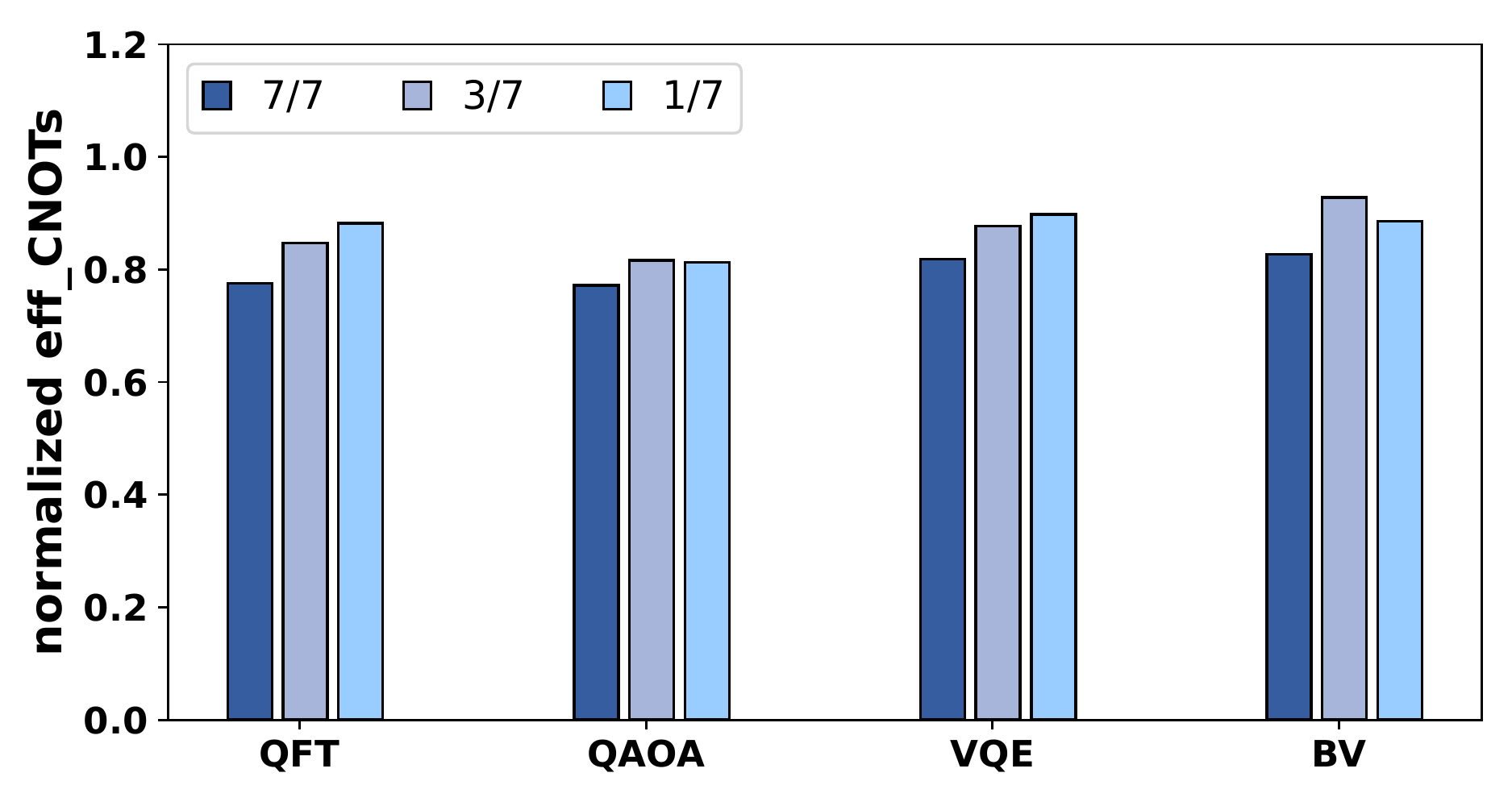}\\
        \vspace{-5pt}(b)
        \caption{Performance of compiled circuits normalized by that of the baseline approach for chiplets with different sparsity levels of cross-chip connections}
        
        \label{fig:cross_chip_sparsity}
\end{figure}
To investigate the impact of sparse cross-chip connections in the chiplet architecture, we evaluate on a 3x3 array of 7x7 chiplets with cross-chip structures at varying sparsity levels. The sparsity was manipulated by keeping 7, 3 and 1 out of the 7 possible cross-chip connections at each edge of a chiplet, denoted as sparsity$=7/7, \enspace 3/7$ and $1/7$, respectively.
Fig.~\ref{fig:cross_chip_sparsity} shows the performance of \frameworkname, with the circuit depth (\ref{fig:cross_chip_sparsity}(a)) and the number of effective CNOTs (\ref{fig:cross_chip_sparsity}(b)) normalized by those of the baseline. It can be seen that as the structures become sparser, the normalized circuit depth decreases, indicating an increased improvement in the circuit depth, while the normalized effective number of CNOTs increases, indicating a decreased improvement in the effective number of CNOTs.
These trends primarily stem from the baseline approach, since \frameworkname\ is relatively stable as the sparsity varies. 
This is because in our framework, cross-chip communications among qubits are achieved primarily by the utilization of highway, whose formation only requires the connectivity among chiplets, but not sensitive to the density of those connections.

\begin{figure}[h!]
        \centering
        \hspace{-15pt}\includegraphics[width=0.86\linewidth]{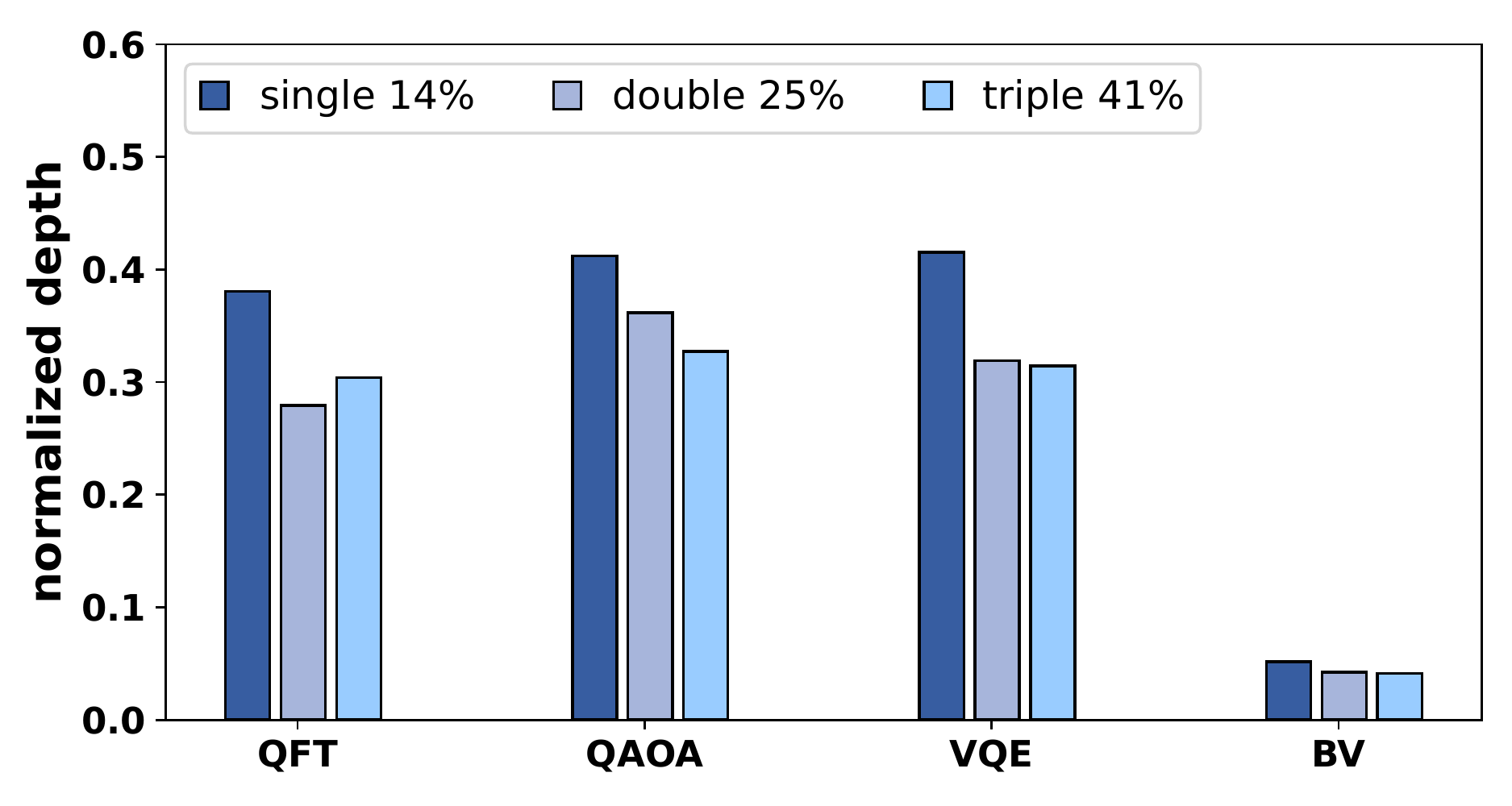} \\
        \vspace{-5pt}(a)\\
        \hspace{-15pt}\includegraphics[width=0.86\linewidth]{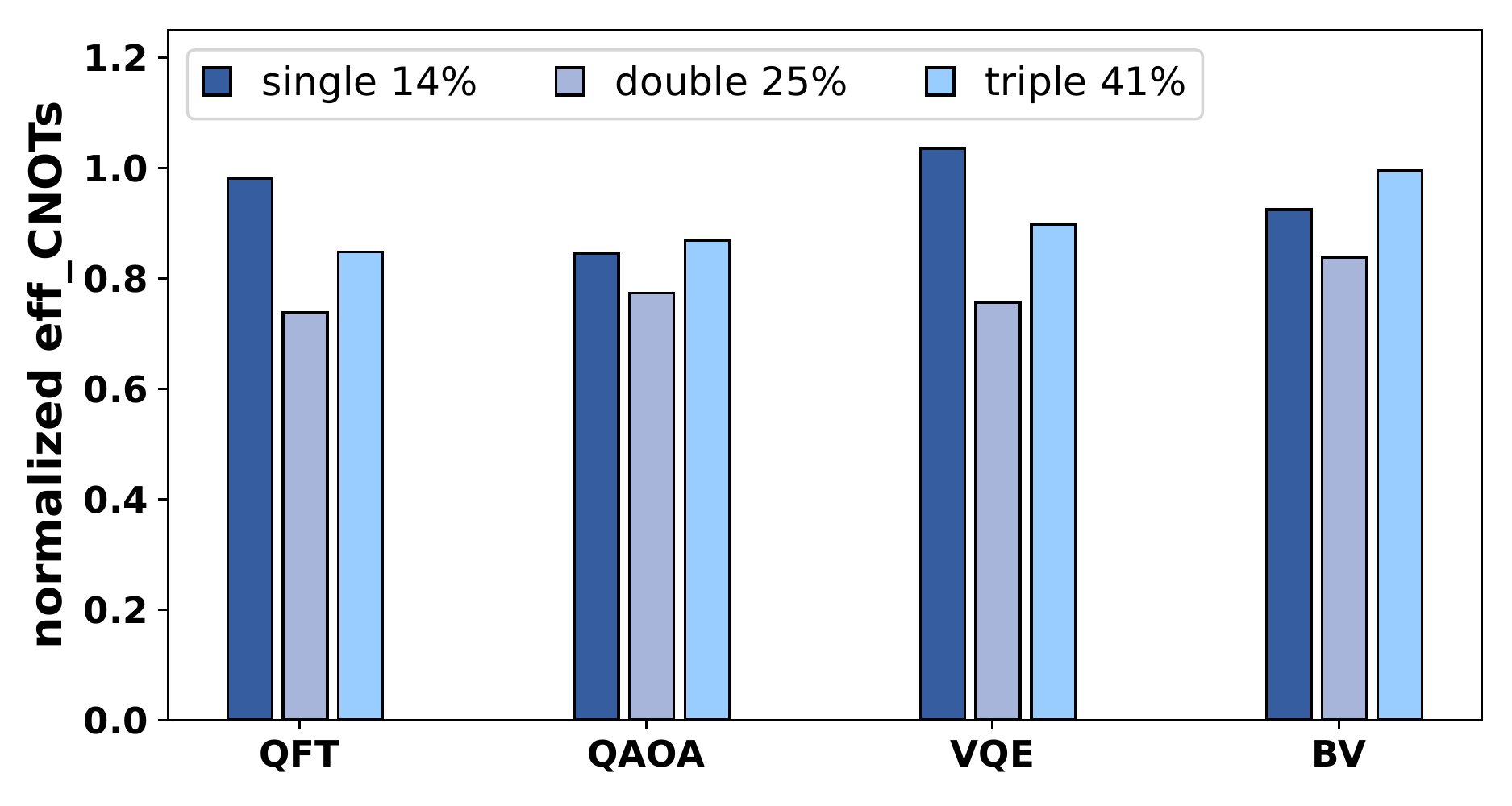}\\
        \vspace{-5pt}(b)
    
        \caption{Performance of compiled circuits normalized by that of the baseline approach for highways with different percentages of highway qubits.}
        
        \label{fig:bandwidth}
\end{figure}

\paragraph{Highway Qubit Percentage}
We explore the effect of highway qubit percentage with a 2x3 array of 9x9 chiplets when the highway paths are similar to Fig.~\ref{fig:highway_configuration_example} (percentage 14\%), doubled (percentage 25\%) and tripled (percentage 41\%), keeping the circuit sizes of the baseline the same as the number of data qubtis. For comparison, the circuit depth (Fig.~\ref{fig:bandwidth}(a)) and the number of effective CNOTs (Fig.~\ref{fig:bandwidth}(b)) are normalized by those of the baseline. It can be seen that the normalized depth is reduced when the highway qubit percentage increases from 14\% to 25\%, but is not affected much as the percentage increases further. The normalized effective CNOT also drops as the highway qubit percentage increases to 25\%, but then slightly increases when it further increases to 41\%. This is because the increased highway qubits incurs a larger overhead of entanglement generation.

\begin{figure}[h!]
        \centering
        \hspace{-15pt}\includegraphics[width=0.86\linewidth]{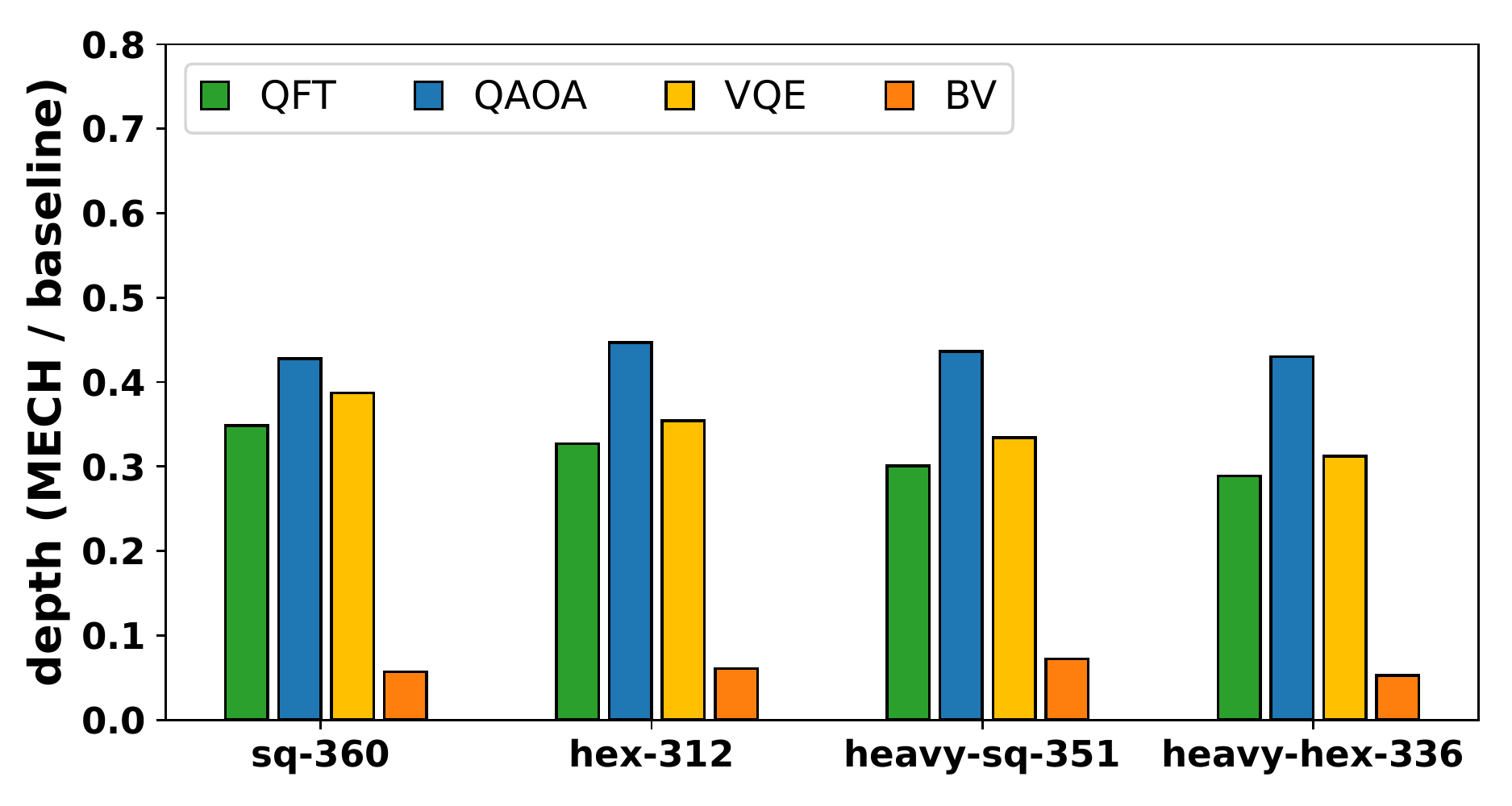} \\
        \vspace{-5pt}(a)\\
        \vspace{5pt}
        \hspace{-15pt}\includegraphics[width=0.86\linewidth]{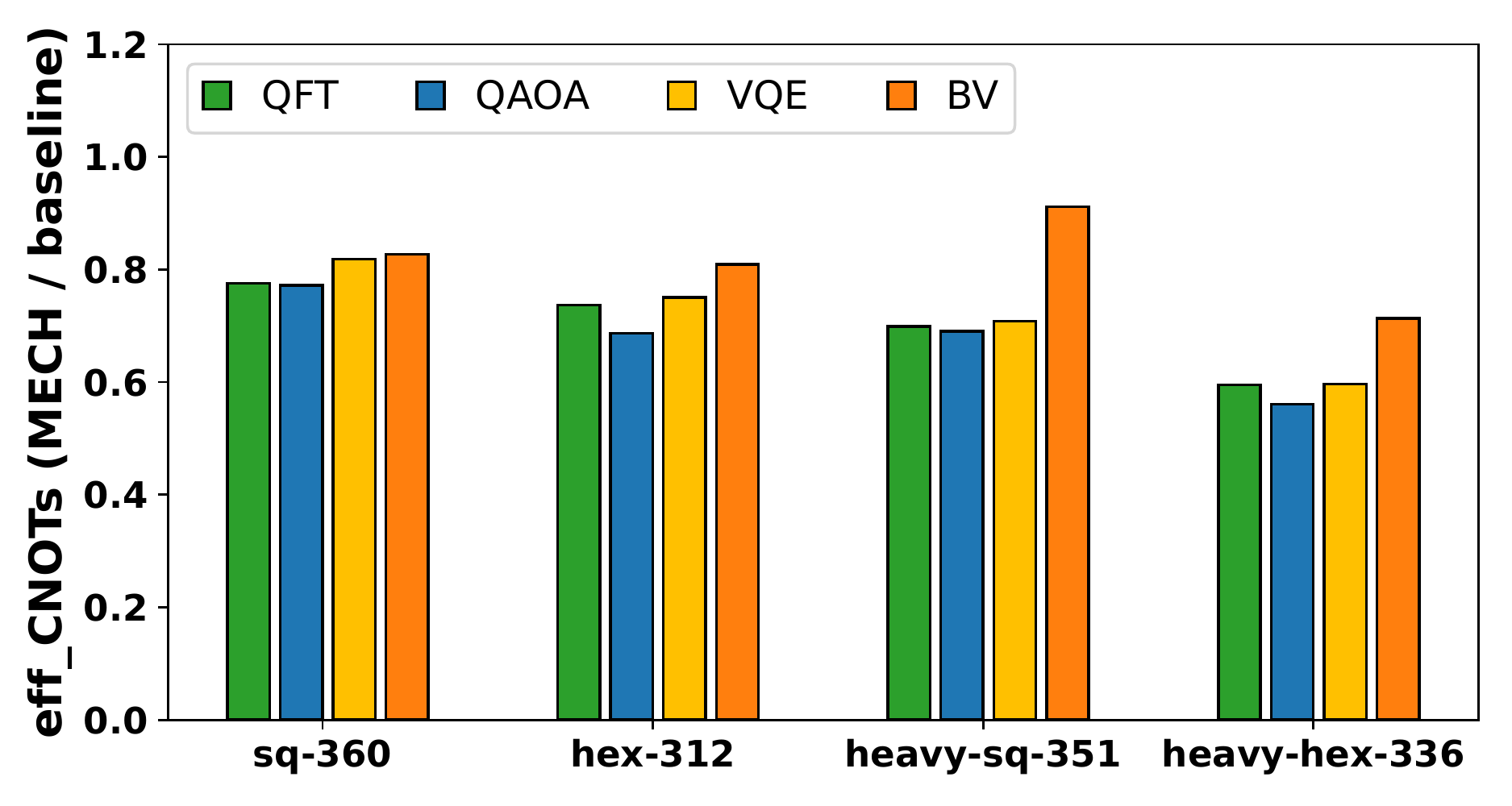}\\
        \vspace{-5pt}(b)
    
        \caption{Performance of compiled circuits normalized by that of the baseline approach for chiplets with various coupling structures.}
        
        \label{fig:different_geometries}
\end{figure}
\paragraph{Coupling Structure}
To demonstrate the generality of \frameworkname, Fig.~\ref{fig:different_geometries} shows the performance of \frameworkname\ on chiplets when the coupling structures are square, hexagon, heavy square and heavy hexagon. For comparison, the circuit depth (\ref{fig:different_geometries}(a)) and the number of effective CNOTs (\ref{fig:different_geometries}(b)) are normalized by those of the baseline. It can be seen that \frameworkname\ achieves similar levels of improvements in performance for all of these structures.
This implies the appilcability of \frameworkname\ for general coupling structures.

\vspace{0.3em}
\section{Conclusion}
In this work, we provide in-depth analysis and discussion of the compilation challenges of scaling up quantum computing with superconducting chiplet architecture. 
We propose a hybrid model of gate-based and measurement-based computing to increase program concurrency with ancillary qubits, and propose a compilation framework of multi-entry communication highway to manage the ancillary resources, facilitating the quantum computing at a larger scale.
We hope that our work could attract more effort from the computer architecture and compiler community to further explore the chiplet architecture and overcome the challenges in the scaling up. 

\section{Acknowledgement}
We thank the anonymous reviewers for their constructive feedback and the cloud bank~\cite{cloud_credit}.
This work is supported in part by Cisco Research, NSF 2048144, NSF 2138437 and Robert N.Noyce Trust.

%
%
%
%
%






\appendix
\section{Artifact Appendix}

\subsection{Abstract}

The artifact contains the source code of MECH compiler and other necessary code scripts to reproduce the key results (Table~\ref{tab:evaluation}, Fig.~\ref{fig:different_chiplet_sizes},\ref{fig:meas_sensitivity},\ref{fig:cross_chip_sparsity},\ref{fig:bandwidth},\ref{fig:different_geometries}) and compare with the baselines in our evaluation. The hardware requirement is a regular X86 server/laptop.
The software dependencies only contain common python packages. We also provide a script to generate the benchmarks used in our evaluation. Note that in the evaluation, results are averaged over multiple executions. So slight deviation is expected in the reproduction.

\subsection{Artifact check-list (meta-information)}


{\small
\begin{itemize}
  \item {\bf Algorithm: } MECH has four core algorithms.
  \begin{itemize}
      \item Circuit.py constructs quantum circuits with gates and measurements, allowing gate commutation to find the earliest execution time of each gate.
      \item Chiplet.py generates chiplet arrays in various geometries (square, hexagon, heavy-square and heavy-hexagon) along with their highway layouts.
      \item HighwayOccupancy.py implements highway routing and dynamic scheduling of highway gates, with the generation and consumption of GHZ states managed automatically.
      \item Router.py further implements local routing, eventually accomplishing the execution of multi-target gates.
  \end{itemize}
  \item {\bf Output: } The output of the compilation is a quantum circuit containing gates and measurements.
  \item {\bf Run-time environment: } Python, Jupyter Notebook.
  \item {\bf Hardware: } Intel CPU, memory size depending on the benchmark size (the largest benchmarks can be processed with 32 GB RAM).
  \item {\bf Disk space required: } 32 GB is sufficient for the artifact and all software dependencies.
  \item {\bf Metrics: } Depth and \#eff\_CNOTs (explained in Section~\ref{sect:eval})
  
  \item {\bf Experiments: } Compiling the benchmark programs with MECH, using Qiskit compiler as the baseline.
  
  \item {\bf Time to complete experiments: } The approximate execution time for each benchmark ranges from 2 minutes -- 10 minutes for MECH, and ranges from 0.5 hour -- 2 hours for the baseline, depending on the benchmark size. It will take hundreds of CPU hours to fully reproduce all results in Table~\ref{tab:evaluation} and Fig.~\ref{fig:different_chiplet_sizes},\ref{fig:meas_sensitivity},\ref{fig:cross_chip_sparsity},\ref{fig:bandwidth},\ref{fig:different_geometries}. However, you can downgrade the optimization level of Qiskit from level 3 to level 2 to obtain similar baseline results much faster.
  \item {\bf Publicly available: } Yes.
  \item {\bf Code licenses: } Apache License 2.0
  \item {\bf Workflow framework used: } Jupyter notebook, Qiskit.
  \item {\bf Archived repo:} https://zenodo.org/record/10544117
  \item {\bf DOI:} 10.5281/zenodo.10544117
\end{itemize}
}

\subsection{Description}

\subsubsection{How to access}

This artifact can be downloaded at the following link \url{https://zenodo.org/record/10544117}.

\subsubsection{Hardware dependencies} A regular server with Intel CPUs can run our artifact while the amount of RAM may limit the size of benchmarks that can be executed. In our experiments, we used 32 GB RAM to execute all benchmarks. If level-3 optimization of Qiskit takes too much memory or time to reproduce the baseline data, level-2 optimization can achieve similar results.

\subsubsection{Software dependencies} The artifact is implemented in Python 3.9.6. Jupyter notebook is needed since we prepared files that contain scripts to automatically and interactively reproduce the results for easy validation. Other dependencies including numpy, matplotlib and networkx. 



\subsection{Installation}
To use our artifact, you may download the repo to your local machine from \url{https://zenodo.org/record/10544117}
and install the software dependencies
by running the command:

\vspace{5pt}
\texttt{pip3 install -r requirements.txt}


\subsection{Evaluation and expected results}

After downloading the artifact and installing all software dependencies, you can open the following jupyter notebook files to reproduce experimental data of baseline and MECH for corresponding table and figures.
\begin{itemize}
    \item Qiskit\_experiements.ipynb (Table~\ref{tab:evaluation}, Fig.~\ref{fig:different_chiplet_sizes},\ref{fig:meas_sensitivity},\ref{fig:cross_chip_sparsity},\ref{fig:bandwidth},\ref{fig:different_geometries}) 
    \item MECH\_experiemnts.ipynb (Table~\ref{tab:evaluation}, Fig.~\ref{fig:different_chiplet_sizes},\ref{fig:meas_sensitivity},\ref{fig:cross_chip_sparsity},\ref{fig:different_geometries})
    \item MECH\_experiemnts\_customized.ipynb (Fig.~\ref{fig:bandwidth})
\end{itemize}

The experimental data will be automatically saved to \texttt{./baseline\_data} and \texttt{./exp\_data}. Then you can open the following jupyter notebook files to read from the saved data and generate the table and plots automatically.

\begin{itemize}
    \item MECH\_main\_table.ipynb (Table~\ref{tab:evaluation})
    \item  MECH\_line\_plot.ipynb (Fig.~\ref{fig:different_chiplet_sizes}, \ref{fig:meas_sensitivity})
    \item MECH\_bar\_plot.ipynb (\ref{fig:cross_chip_sparsity},\ref{fig:bandwidth},\ref{fig:different_geometries})
\end{itemize}
Please refer to README.md for more detailed instructions.








\bibliographystyle{unsrturl}
\bibliography{references}


\end{document}